\documentclass[a4paper,onecolumn,unpublished]{quantumarticle}

\pdfoutput=1

\usepackage{amsmath,amssymb,amsthm}
\usepackage{graphicx}
\usepackage{bm}
\usepackage{booktabs}
\usepackage{xcolor}

\usepackage[numbers,sort&compress]{natbib}
\usepackage{hyperref}

\newtheorem{theorem}{Theorem}
\newtheorem{lemma}{Lemma}
\newtheorem{corollary}{Corollary}
\newtheorem{proposition}{Proposition}
\theoremstyle{definition}
\newtheorem{stdresult}{Standard Result}[section]

\newcommand{\Msf}{\mathcal{M}}
\newcommand{\Psf}{\mathcal{P}}
\newcommand{\Ncov}{N_{\mathrm{cov}}}
\newcommand{\E}{\mathbb{E}}
\newcommand{\Prob}{\mathbb{P}}
\DeclareMathOperator{\Bin}{Bin}
\newcommand{\covers}{\succeq}
\newcommand{\pmax}{p_{\max}}
\newcommand{\zd}{z_\delta}
\newcommand{\wstar}{w^\star}
\newcommand{\He}{\mathrm{He}}
\newcommand{\epslat}{\varepsilon_{\mathrm{lat}}}
\newcommand{\cbar}{\bar c}
\newcommand{\nueff}{\nu}
\newcommand{\stdref}[1]{Standard Result~\ref{#1}}
\newcommand{\KL}{D_{\mathrm{KL}}}
\newcommand{\Wal}{\mathsf{H}}
\newcommand{\Nsh}{N_{\mathrm{sh}}}
\newcommand{\Mobs}{M_{\mathrm{obs}}}
\newcommand{\phat}{\hat p}

\begin{document}

\title{Covering the Pauli Observables: Sampling Thresholds and the Design
Effect of a Measurement Setting}

\author{Kausthubh Chandramouli}
\affiliation{Department of Electrical and Computer Engineering, North Carolina State University, Raleigh, NC 27606, USA}
\email{kprabha@ncsu.edu}

\author{Noah Siekierski}
\affiliation{Department of Mathematics, North Carolina State University, Raleigh, NC 27695, USA}
\affiliation{Department of Electrical and Computer Engineering, North Carolina State University, Raleigh, NC 27606, USA}
\affiliation{Pritzker School of Molecular Engineering, University of Chicago, Chicago, IL 60637, USA}

\author{Bojko N. Bakalov}
\affiliation{Department of Mathematics, North Carolina State University, Raleigh, NC 27695, USA}

\author{Dror Baron}
\affiliation{Department of Electrical and Computer Engineering, North Carolina State University, Raleigh, NC 27606, USA}

\maketitle

\begin{abstract}
Quantum state tomography and shadow-based estimation both rely on measuring Pauli observables.
In practice, an experiment does not measure these observables one at a time. It chooses a
measurement setting that assigns one of the bases $X$, $Y$ or $Z$ to each qubit, and a single
run in that setting yields estimates of every Pauli observable compatible with it. We say that
the setting \emph{covers} those observables. We determine how many randomly chosen settings are
needed to cover a prescribed fraction of all Pauli observables. Since the number of settings
grows exponentially with the number of qubits, the covered fraction has a sharp threshold at a
growth rate of $3^{3/4}$ per qubit. Covering is an instance of lossy data compression at zero distortion,
and the threshold is set by the rate--distortion function between an observable and a collection of settings that cover it. We give an explicit expression
for the number of settings needed to reach a target fraction, accurate up to a remainder that
reduces to a small oscillation for many qubits. We also show that random sampling
is asymptotically optimal in the number of settings, up to a factor of less than $1.69$.
However, coverage does not determine what a setting contributes to an
estimator. All observables covered by a setting are computed from the same measurement
outcomes, so their estimates are correlated. We quantify this correlation through the design
effect of a setting, which equals the sum of the squared Pauli expectations that the setting
captures and depends on the target state. We derive its mean over settings in closed form for every
state, evaluate it for the Greenberger--Horne--Zeilinger (GHZ) and W states and on average over
Haar-random states, and show that it can be estimated without bias from the measurement data alone.
In simulations on six qubits, measuring settings needs $7.6$ to $38$ times fewer circuit runs than
measuring the same number of observables independently. At high coverage, fitting the outcome
histograms instead of averaged estimates roughly halves the reconstruction error, mainly because the
fit counts how many settings cover each observable. The correlation itself costs such a fit about a
factor of two, and a larger design effect signals more to gain from modeling it.
\end{abstract}

\section{Introduction}

Measurement of Pauli observables is a basic step in quantum state tomography and in shadow-based
estimation \cite{cramer2010efficient,gross2010compressed,siekierski2025approximate,huang2020predicting}.
On $n$ qubits, a Pauli observable is a tensor product $P=P_1\otimes\cdots\otimes P_n$ with each
$P_i\in\{I,X,Y,Z\}$, so there are $4^n$ of them. An experiment does not measure these observables
one at a time. Before each run, it chooses one of the bases $X$, $Y$ or $Z$ for every qubit, a
choice that we call a \emph{measurement setting}. The run returns one bit per qubit, and every Pauli
observable that agrees with the setting on the qubits where it is not the identity can be
estimated from that bit-string by multiplying the relevant bits. We say that the setting
\emph{covers} these observables. Each of the $3^n$ settings covers $2^n$ observables, and much of
the work on classical shadows and derandomized Pauli estimation
\cite{huang2020predicting,huang2021efficient,elben2023toolbox,hadfield2022measurements,gresch2025guaranteed,reggio2024fast}
is concerned with choosing settings that make many observables estimable.

An experimentalist therefore has to decide, before collecting any data, how many settings to schedule,
and needs to know what fraction of the observables those settings will cover. For a single
observable, the answer is well known: a uniformly random setting covers a fixed observable of
weight $w$, that is, one acting nontrivially on $w$ qubits, with probability $3^{-w}$. This is the reason
the sample complexity of classical shadows carries a factor $3^{w}$ for Pauli observables of weight $w$
\cite{huang2020predicting}. For a whole collection of random settings, the growth rate of the
number of settings needed follows from the rate--distortion theorem (Section~\ref{ss:prior}). The
higher-order terms, which determine how many settings an experiment should use, have not been worked
out for Pauli measurements, to our knowledge. Covering \emph{all} $4^n$ observables requires all $3^n$
settings, since each weight-$n$ observable is covered only by the setting that coincides with it,
and dropping any one setting destroys informational completeness. We
therefore ask how many random settings are needed to cover a fraction $1-\delta$ of the
observables, for a fixed $\delta\in(0,1)$.

Our first result is a sharp threshold. If the number of settings grows as $\beta^n$, the covered
fraction tends to one when $\beta>3^{3/4}\approx2.28$ and to zero when $\beta<3^{3/4}$, even
for the best choice of settings (Theorem~\ref{thm:threshold}). The rate $3^{3/4}$ is determined by the bulk of the weight distribution:
most observables have a weight close to $\tfrac34n$, and a random setting covers each of them with
a probability about $3^{-3n/4}$. One of our key results is an explicit expression for the number of settings required to reach a target coverage fraction. Writing
$k^\star(n,\delta)$ for the least number of random settings whose expected coverage reaches a
fraction $1-\delta$, we show (Theorem~\ref{thm:kdelta}) that
\begin{equation}\label{eq:kstar-headline}
   \log_3 k^\star(n,\delta)=\frac{3n}{4}+\frac{\sqrt{3n}}{4}\,\Phi^{-1}(1-\delta)
      -\frac{\Phi^{-1}(1-\delta)^2-1}{12}-\frac{\gamma}{\ln3}+\frac{c(\delta)}{\sqrt n}
      +\varepsilon_n(\delta),
\end{equation}
where $\Phi$ is the standard normal cumulative distribution function, $\gamma\approx0.5772$ is
the Euler--Mascheroni constant, and $c(\delta)$ is a cubic polynomial in $\Phi^{-1}(1-\delta)$
given in Theorem~\ref{thm:kdelta}. The $\sqrt{3n}$ term comes from the Gaussian fluctuation of
the weight, and the term in $\Phi^{-1}(1-\delta)^2$ from its skewness. The last two explicit terms
come mostly from the gradual transition between the weights that are covered and those that are
not. This transition shifts the crossover weight by its mean, which gives $-\gamma/\ln3$, and
widens the weight distribution by its variance, which gives most of $c(\delta)/\sqrt n$. For large
$n$, the remainder $\varepsilon_n(\delta)$ is at most $2.4\times10^{-4}$ in magnitude, and its
leading part is an oscillation that is periodic in $\log_3k^\star$. At practical sizes, the
remainder decays like $1/n$. For $n\ge80$ the explicit terms are within $2\%$ of $k^\star$ at
$\delta=0.1$ and within $10\%$ at $\delta=10^{-3}$. Random sampling is
also close to optimal: for large $n$, random sampling needs fewer than $1.69$ times as many settings as the best selection rule
(Theorem~\ref{thm:lb} and Corollary~\ref{cor:constant}).
When only observables of weight at most $\omega n$ matter, for a fixed $\omega<\tfrac34$, the rate
drops to $3^{\omega}$ and the
$\sqrt n$ term disappears (Corollary~\ref{cor:bounded}).

Coverage treats all settings alike, but an estimator does not. The $2^n$ observables covered by a setting
are all computed by marginalizing the same outcome distribution, so their estimates are
correlated. We describe this correlation in Section~\ref{sec:design} through the effective support size
$\nueff=(\sum_bp(b)^2)^{-1}$ of the outcome distribution $p$ of the setting on
the target state.
We call $D=2^n/\nueff$ the \emph{design effect} of the setting, after the
corresponding quantity in survey sampling \cite{kish1965survey}. It equals the sum of the squared
expectations of the observables that the setting covers (Proposition~\ref{prop:mass}), and $D-1$
measures how far the noise of the setting is from white. The design effect depends on the
state. Averaged over Haar-random states, its mean over settings is $2\cdot2^n/(2^n+1)<2$, and it
grows like $\tfrac12(4/3)^n$ for the Greenberger--Horne--Zeilinger (GHZ) state and the W state
(Proposition~\ref{prop:design} defines both states and gives these values). The design effect can
be estimated without bias from the counts the experiment already records. We simulate six-qubit
tomography with Riemannian gradient descent and approximate message passing \cite{hsu2024quantum,siekierski2025approximate}. At $93\%$ coverage,
fitting the outcome histograms instead of averaged estimates roughly halves the reconstruction
error, because the fit weights each observable by the number of settings that cover it. The
correlation itself costs such a fit about a factor of two, which a maximum-likelihood reference
recovers, and the reference gains more for states with a larger mean design effect. A histogram fit
reaches an infidelity of $3\times10^{-4}$ with $24$ to $76$ times fewer circuit runs than measuring
observables independently. Much of that saving is the ratio of the number of observables to the
number of settings, $7.6$ to $38$ depending on coverage, and every saving is measured against the
independent-observable model of compressed-sensing and message-passing analyses
\cite{gross2010compressed,siekierski2025approximate}.

\subsection{Relation to prior work}\label{ss:prior}

Choosing the fewest settings that cover a prescribed collection of observables is a set-cover
problem with a ground set of $4^n$ observables and $3^n$ sets of size $2^n$. The classical analysis of the greedy heuristic
\cite{johnson1974approximation,lovasz1975ratio,chvatal1979greedy,slavik1997tight} bounds the gap
between greedy and optimal selection by a factor $O(\ln2^n)=O(n)$, and the hardness result of Feige
\cite{feige1998threshold} shows that, under standard complexity assumptions, no polynomial-time
rule improves this factor on general set systems. That factor adds $O(\log n)$ to $\log_3k$, which leaves the
rate $3^{3/4}$ and the $\sqrt n$ term of \eqref{eq:kstar-headline} unchanged but is larger than the
constants we compute.

Covering has been studied in a more general form in information theory. It is an instance of
lossy source coding at zero distortion, and the ternary analogue of the binary erasure quantization
of Martinian and Yedidia \cite{martinian2003iterative}, in which a reproduction may set an erased
source bit to either value. The threshold $3^{3/4}$ is the exponential of the rate--distortion
function (Section~\ref{ss:threshold}). The $\sqrt{3n}$ term of \eqref{eq:kstar-headline} has the
form of the dispersion term, the $\sqrt n$ correction to the rate that Kostina and Verd\'u
\cite{kostina2012fixed}, and independently Ingber and Kochman \cite{ingber2011dispersion}, found for finite-blocklength lossy compression at
positive distortion (Section~\ref{ss:opt}). The constants $\gamma/\ln3$ and $\epslat$ of
Theorem~\ref{thm:kdelta} are classical constants from the expression for the maximum of independent geometric random
variables \cite{szpankowski1990yet,eisenberg2008expectation,louchard2006asymptotics}
(Section~\ref{ss:cost}).

Sampling settings rather than observables is what product-basis experiments do. Riofr\'io et al.\
\cite{riofrio2017experimental} measured $127$ of the $3^7$ Pauli settings in a seven-qubit
compressed-sensing experiment, and Steffens et al.\ \cite{steffens2017experimentally} studied
reconstruction quality against the number of randomly chosen settings. Siekierski et al.\
\cite{siekierski2025approximate} tabulate, for up to ten qubits, simulated numbers of random settings
that cover prescribed fractions of the observables. None of these works gives a rule for
how many settings a prescribed coverage needs. Qin et al.\ \cite{qin2024optimal} analyze the
trade-off between the number of Pauli observables measured and the repetitions of each, for
independently measured observables. Goldar et al.\ \cite{goldar2026exponential} use the $3^n$ product bases
as the measurement model of an adaptive-tomography separation, and Keenan, Goold and Nico-Katz
\cite{keenan2025random} analyze the protocol that measures all $3^n$ settings. Requiring
\emph{every} observable of weight at most $w$ to be covered is the construction of a covering array
\cite{colbourn2004combinatorial,sarkar2017upper}, and its quantum form, overlapping tomography
\cite{cotler2020overlapping,evans2019scalable,bonetmonroig2020nearly,hansenne2025optimal,wei2025optimal},
needs an extra factor logarithmic in $n$ for that guarantee.
Asking for a fraction of the observables avoids the extra factor.

Locally biased classical shadows \cite{hadfield2022measurements}, derandomization
\cite{huang2021efficient} and ShadowGrouping \cite{gresch2025guaranteed} choose settings for a
prescribed set of target observables. The covered fraction of all $4^n$ observables is not their
objective. The covariance between Pauli observables read from the same setting, which
Section~\ref{sec:design} builds on, is also known. Crawford et al.\ \cite{crawford2021efficient}
use the covariance to allocate shots across commuting groups, Yen, Ganeshram and Izmaylov
\cite{yen2023deterministic} build grouping schemes on covariance estimates, Wu et al.\
\cite{wu2023overlapped} study the overlap structure of grouped measurements, and Verteletskyi, Yen
and Izmaylov \cite{verteletskyi2020measurement} frame grouping as a minimum clique cover. Keenan,
Goold and Nico-Katz \cite{keenan2025random} study this covariance and its dependence on the state
when all settings are measured, Acharya, Kypraios and Gu\c{t}\u{a} \cite{acharya2019comparative}
compare tomography estimators that include generalized least squares with the multinomial
covariance of each setting, and Hayashi \cite{hayashi2026finite} gives finite-sample bounds on compressions of
the classical-shadow covariance.

Our contribution is to bring the information-theoretic results above to Pauli measurement
settings and to connect the sampling of settings to estimation. On the covering side,
this connection provides the explicit count \eqref{eq:kstar-headline}, the $\sqrt{3n}$ term at
zero distortion with an $O(1)$ remainder and a comparison of random sampling with any selection
rule. On the estimation side, the design effect
links a sampled setting to the estimators that use its data. We give its Walsh--Hadamard form, its
mean over settings for every state and an unbiased estimate from the counts, and we measure its
cost to a reconstruction algorithm in simulation. Section~\ref{sec:setup} fixes notation and
computes the expected coverage, Section~\ref{sec:covering} gives the covering results,
Section~\ref{sec:design} studies the design effect, Section~\ref{sec:numerics} reports the
simulations, and Section~\ref{sec:conclusion} lists open questions.

\section{Setup and expected coverage}\label{sec:setup}

In this section, we fix the notation for observables, settings and their weights, and we show
that the weight of a uniformly drawn observable is binomial. We then compute the expected number of
observables that $k$ random settings cover, with and without replacement, and show that the two
agree to leading order.

\subsection{Notation}\label{ss:notation}
Write $[d]=\{1,\dots,d\}$. A \emph{Pauli observable} is a string
$P=(P_1,\dots,P_n)\in\{I,X,Y,Z\}^n=:\Psf_n$ and a \emph{measurement setting} is a string
$M=(M_1,\dots,M_n)\in\{X,Y,Z\}^n=:\Msf_n$, so $|\Psf_n|=4^n$ and $|\Msf_n|=3^n$. We use the same symbol for the string $P$ and for the operator
$P_1\otimes\cdots\otimes P_n$ that it labels, where $I$ is the $2\times2$ identity and $X$, $Y$ and
$Z$ are the Pauli matrices. Likewise, the letter $M_i$ of a setting names the Pauli matrix whose
eigenbasis is measured on qubit $i$. The \emph{weight}
of an observable is the number of its non-identity coordinates,
\begin{equation}\label{eq:weight-def}
   |P|:=\#\{i\in[n]:P_i\neq I\},
\end{equation}
and the observables of weight $w$ form the \emph{weight-$w$ shell}. A setting $M$ \emph{covers} an
observable $P$, written $M\covers P$, when $P_i\in\{I,M_i\}$ for every $i\in[n]$. The following
counting facts are used throughout:
\begin{align}
   \#\{M\in\Msf_n: M\covers P\}&=3^{n-|P|},\label{eq:F1}\\
   \#\{P\in\Psf_n: |P|=w\}&=\binom nw 3^{w}.\label{eq:F2}
\end{align}
For \eqref{eq:F1}, a covering setting is forced on the $|P|$ non-identity coordinates and free on
the others. For \eqref{eq:F2}, choose which $w$ coordinates are non-identity and then one of
$X,Y,Z$ on each. Conversely, a fixed setting $M$ covers $2^n$ observables, since coordinates of
covered observables are either $I$ or $M_i$.

For a collection $\mathcal S=\{M^{(1)},\dots,M^{(k)}\}\subseteq\Msf_n$ of $k$ settings, let
\begin{equation}\label{eq:Ncov-def}
   \Ncov(\mathcal S):=\#\bigl\{P\in\Psf_n:\ M^{(j)}\covers P\ \text{for some }j\in[k]\bigr\}
\end{equation}
be the number of covered observables. When $\mathcal S$ is a uniformly random $k$-subset of
$\Msf_n$, that is, under sampling without replacement, we write $\Ncov(k)$. We use the same
notation when the $k$ settings are sampled with replacement. We specify whether settings are sampled with or without replacement throughout the text as needed.

We study sequences of problems in which the number of settings grows exponentially in the number
of qubits, that is, sequences $k=k_n$ for which
\begin{equation}\label{eq:kbeta}
   \beta:=\lim_{n\to\infty}k_n^{1/n}\quad\text{exists, with }\beta\in(1,3].
\end{equation}
We call $\beta$ the \emph{growth base} and write $k=\beta^n$ as shorthand for \eqref{eq:kbeta}.

If an observable is drawn uniformly from $\Psf_n$, each coordinate is identity with probability $\tfrac14$, so by \eqref{eq:F2} its weight $W$ is binomial:
\begin{equation}\label{eq:weightdist}
   W\sim\Bin\bigl(n,\tfrac34\bigr),\qquad p_w:=\Prob[W=w]=\binom{n}{w}3^w 4^{-n},
\end{equation}
with mean and standard deviation
\begin{equation}\label{eq:musigma}
   \mu_W=\tfrac34n,\qquad \sigma_W=\tfrac{\sqrt{3n}}{4}.
\end{equation}
Grouping observables by weight in this way is the method of types
\cite{cover2006elements,csiszar1998method} for the alphabet $\{I,X,Y,Z\}$ with the identity letter
singled out.

We write $\Phi$ for the standard normal cumulative distribution function, $\phi=\Phi'$ for its
density, and
\begin{equation}\label{eq:zdelta}
   \zd:=\Phi^{-1}(1-\delta),
\end{equation}
which is positive for $\delta<\tfrac12$ and negative for $\delta>\tfrac12$. Asymptotic statements
are taken as $n\to\infty$ with $\delta\in(0,1)$ fixed, so the constants hidden in $O(\cdot)$ and
$o(\cdot)$ may depend on $\delta$ but not on $n$. We write
\begin{equation}\label{eq:dap}
\KL(a\,\|\,p)=a\ln\frac ap+(1-a)\ln\frac{1-a}{1-p}
\end{equation}
for the binary relative entropy and $\gamma\approx0.5772$ for the Euler--Mascheroni constant.

\subsection{Expected coverage}\label{ss:expected}
\begin{lemma}[Expected coverage]\label{lem:exp}
Under uniform sampling without replacement of $k\le3^n$ settings, we have
\begin{equation}\label{eq:Ecov}
   \E[\Ncov(k)]=4^n\Bigl(1-\sum_{w=0}^{n}p_w\,q_w(k)\Bigr),\qquad
   q_w(k)=\binom{3^n-3^{n-w}}{k} \binom{3^n}{k}^{\!\!-1},
\end{equation}
where $q_w(k)$ is the probability that a fixed observable of weight $w$ is left uncovered.
Moreover,
\begin{equation}\label{eq:sandwich}
   \Bigl(1-\frac{3^{-w}}{1-k3^{-n}}\Bigr)^{k}\ \le\ q_w(k)\ \le\ (1-3^{-w})^{k},
\end{equation}
where the upper bound holds for every $w$ and the lower bound holds whenever
$k\le3^n-3^{n-w}$.
\end{lemma}

\begin{proof}
By linearity of expectation, $\E[\Ncov(k)]=4^n-\sum_P\Prob[P\text{ uncovered}]$. By
\eqref{eq:F1}, an observable of weight $w$ is covered by $3^{n-w}$ of the $3^n$ settings, so it is
uncovered if and only if all $k$ sampled settings fall among the other $3^n-3^{n-w}$, which has
probability $q_w(k)$. Grouping observables by weight with \eqref{eq:F2} gives \eqref{eq:Ecov}. For
the bounds, write $q_w(k)$ as a product of conditional probabilities,
\begin{equation}\label{eq:telescope}
   q_w(k)=\prod_{i=0}^{k-1}\Bigl(1-\frac{3^{-w}}{1-i\,3^{-n}}\Bigr),
\end{equation}
where the factor with index $i$ is the probability that the next draw misses the observable given
that the previous $i$ draws did. The factors lie in $[0,1]$ and decrease with $i$. Bounding each
by the factor at $i=0$ gives the upper bound. Each factor is also at least
$1-3^{-w}(1-k3^{-n})^{-1}$, which is nonnegative when $k\le3^n-3^{n-w}$, and multiplying these
lower bounds gives the lower bound.
\end{proof}

The upper bound in \eqref{eq:sandwich} is the probability of missing the observable when the $k$
settings are drawn independently, and the lower bound differs from it only through the factor
$(1-k3^{-n})^{-1}$, which tends to one when $\beta<3$. Sampling with and without replacement
therefore agree to leading order. Sampling without replacement saves $33\%$ of the settings at
$n=6$, $12\%$ at $n=12$ and $2.5\%$ at $n=20$ for $\delta=0.1$, and less than $0.01\%$ by $n=50$.
It is convenient to work with the uncovered fraction under sampling with replacement,
\begin{equation}\label{eq:Udef}
   U(k):=\sum_{w=0}^{n}p_w\,(1-3^{-w})^{k},
\end{equation}
its counterpart without replacement $\widetilde U(k):=\sum_w p_wq_w(k)$, and the lower bound
\begin{equation}\label{eq:Ulower}
   \underline U(k):=\sum_{w=1}^n p_w\Bigl(1-\frac{3^{-w}}{1-k3^{-n}}\Bigr)^{k}.
\end{equation}
Lemma~\ref{lem:exp} supplies the lower bound in \eqref{eq:sandwich} only when
$k\le3^n-3^{n-w}$, which fails at $w=0$. We can safely omit the $w=0$ term: the only weight-$0$
observable is the identity $I^{\otimes n}$, which every setting covers, so $q_0(k)=0$ for every
$k\ge1$.  For every $w\ge1$, the
hypothesis holds once $n$ is large, because $k3^{-n}\to0$ when $\beta<3$. Hence
$\underline U(k)\le\widetilde U(k)\le U(k)$ for large $n$, and
$\E[\Ncov(k)]/4^n=1-\widetilde U(k)$.

\section{Covering results}\label{sec:covering}

In this section, we find how many random settings cover a prescribed fraction of the observables,
first at the level of the growth rate and then up to a bounded remainder. We then compare random
sampling with an optimal selection strategy. Finally, we restrict attention to observables of
bounded weight, for which fewer settings suffice.

Our covering results come from comparing the counting fact
\eqref{eq:F1} with the weight law \eqref{eq:weightdist}. The miss probability $(1-3^{-w})^k\approx\exp(-k3^{-w})$ changes from
nearly zero to nearly one as $w$ passes the \emph{crossover weight}
\begin{equation}\label{eq:wstar}
   \wstar:=\log_3 k,
\end{equation}
so $k$ settings cover almost every observable of weight well below $\wstar$ and almost none of
weight well above it. The covered fraction is therefore governed by where $\wstar$ falls in the
weight distribution \eqref{eq:weightdist}.

\subsection{A sharp threshold}\label{ss:threshold}

The covering problem is an instance of lossy source coding \cite[Ch.~10]{cover2006elements},
that is, compression in which the reproduction may differ from the source within a prescribed
distortion. The source emits an observable drawn uniformly from $\Psf_n$, a code is a collection of
settings, and the distortion between $P$ and $M$ is $\sum_id(P_i,M_i)$ with
\begin{equation}\label{eq:dontcare}
   d(P_i,M_i)=\begin{cases}0, & P_i=I\ \text{or}\ P_i=M_i,\\ 1, & \text{otherwise.}\end{cases}
\end{equation}
A setting therefore covers an observable if and only if it reproduces
the observable with zero distortion. The identity plays the role of the erased symbol in the binary
erasure quantization of Martinian and Yedidia \cite{martinian2003iterative}. The rate--distortion
function at zero distortion, which is the least mutual information per qubit between an observable and a
reproduction that covers it, is
\begin{equation}\label{eq:R0}
   R(0)=\ln4-h\bigl(\tfrac14\bigr)=\tfrac34\ln3\ \text{nats},
\end{equation}
where $h$ is the binary entropy in nats. The minimum is achieved when an identity factor is
reproduced uniformly on $\{X,Y,Z\}$: a uniformly drawn factor $P_i$ has entropy $\ln4$, and given
$M_i$ it is $I$ with probability $\tfrac14$ and $M_i$ otherwise, which leaves the conditional entropy $h(\tfrac14)$.
Zero distortion forces $P_i\in\{I,M_i\}$, so given $M_i$ the factor
$P_i$ is determined by whether it equals $I$, which has probability $\tfrac14$. Hence
$H(P_i\mid M_i)\le h(\tfrac14)$ and $I(P_i;M_i)\ge\ln4-h(\tfrac14)$.
Theorem~\ref{thm:threshold} is the rate--distortion theorem for this distortion, with threshold
$e^{R(0)}=3^{3/4}$. Part~(i) is the covering lemma \cite[Lemma~3.3]{elgamal2011network} applied to
this distortion, with the shells near weight $\tfrac34n$ in the role of the typical set, the set of
source sequences that carries almost all the probability. Part~(ii) is its strong converse: below
the rate, the covered fraction tends to zero for any collection of settings. We include a short proof below
to keep the paper self-contained.

\begin{theorem}[Threshold]\label{thm:threshold}
Suppose that $k=k_n$ satisfies \eqref{eq:kbeta} with growth base $\beta\in(1,3)$.
\begin{enumerate}
\item[\textup{(i)}] If $\beta>3^{3/4}$ and the settings are sampled uniformly, with or without
replacement, then $\E[\Ncov(k)]/4^n\to1$ and $\Ncov(k)/4^n\to1$ in
probability.
\item[\textup{(ii)}] If $\beta<3^{3/4}$, then   $\max_{|\mathcal S|=k}\Ncov(\mathcal S)/4^n\to0$, where the maximum runs over all
collections of $k$ settings.
\end{enumerate}
\end{theorem}

\begin{proof}
{(i)} Without replacement, $\underline U(k)\le\widetilde U(k)\le U(k)$ for large $n$. With replacement,
the uncovered fraction is $U(k)$ itself.  In both cases, it suffices to show that $U(k)\to0$.  Fix $\alpha$ with $\tfrac34<\alpha<\log_3\beta$ and split the sum \eqref{eq:Udef} at
weight $\alpha n$. The observables of weight above $\alpha n$ have a total probability
$\Prob[W>\alpha n]\le e^{-n\KL(\alpha\|3/4)}$ by the Chernoff bound (\stdref{std:chernoff}), which tends to zero. For
$w\le\alpha n$ we have $k3^{-w}\ge3^{cn(1+o(1))}$ with $c=\log_3\beta-\alpha>0$, so by the inequality $(1-x)^k\le e^{-kx}$ (\stdref{std:elem}), we have
\[
   (1-3^{-w})^k\le e^{-k3^{-w}}\le\exp\bigl(-3^{cn(1+o(1))}\bigr),
\]
which is doubly exponentially small. Hence, $U(k)\to0$ and $\E[\Ncov(k)]/4^n\to1$. The uncovered
count $4^n-\Ncov(k)$ is nonnegative with mean $4^n\widetilde U(k)$, and Markov's inequality gives
$\Prob[\Ncov(k)\le(1-\tau)4^n]\le\widetilde U(k)/\tau\to0$ for every $\tau>0$.

{(ii)}   A setting covers an observable of
weight $w$ only if it matches the observable on its $w$ non-identity coordinates, so the observable
is determined by the setting and by the choice of those $w$ coordinates. Hence one setting covers
$\binom nw$ observables of weight $w$, and $k$ settings cover at most $\min(\binom nw3^w,k\binom
nw)$ of them. Summing over $w$, every collection $\mathcal S$ of $k$ settings satisfies the
\emph{counting bound}
\begin{equation}\label{eq:anybound}
   \Ncov(\mathcal S)\ \le\ \sum_{w=0}^{n}\binom{n}{w}\min\bigl(3^w,\,k\bigr).
\end{equation}
Put $\bar w=\lceil\log_3(kn)\rceil$, so that $\bar w/n\to\log_3\beta<\tfrac34$. Dividing
\eqref{eq:anybound} by $4^n$ bounds the covered fraction by $\sum_wp_w\min(1,k3^{-w})$. Each term with $w\ge\bar w$ is at
most $p_wk3^{-w}\le p_w/n$, so these terms sum to at most $1/n$. Hence $\Ncov(\mathcal S)/4^n\le\Prob[W<\bar
w]+1/n$ for every $\mathcal S$, which tends to zero by the law of large numbers because $\bar w/n$
stays below the mean fraction $\tfrac34$.
\end{proof}

The value $\tfrac34$ is the mean weight fraction. Covering a constant fraction of the shell of
weight $\tfrac34n$ already needs about $3^{3n/4}$ settings, because each setting covers a
$3^{-3n/4}$ fraction of that shell, and by \eqref{eq:weightdist} that shell and its neighbors
carry almost all observables.

\begin{figure}[!htb]
\centering
\includegraphics[width=0.6\linewidth]{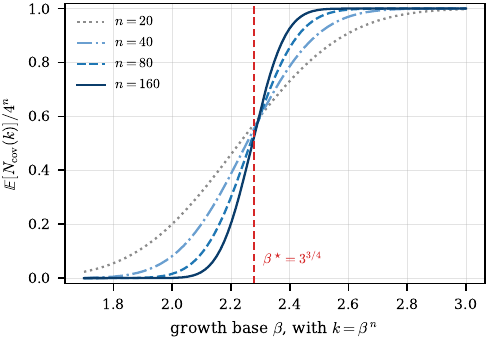}
\caption{Expected covered fraction $\E[\Ncov(k)]/4^n$ under uniform sampling of $k$ settings without
replacement, computed from \eqref{eq:Ecov}, against the growth base $\beta$, with
$k=\beta^{n}$, for $n=20,40,80,160$. The curves steepen into a step at $\beta=3^{3/4}$
(dashed), as Theorem~\ref{thm:threshold} predicts. At $n=160$, the covered fraction is
$7.5\times10^{-4}$ at $\beta=2.0$, $0.53$ at $\beta=3^{3/4}\approx2.28$ and $0.93$ at
$\beta=2.4$.}
\label{fig:threshold}
\end{figure}

Figure~\ref{fig:threshold} shows the threshold at finite $n$. As $n$ grows, the expected covered
fraction $\E[\Ncov(k)]/4^n$ under uniform sampling without replacement steepens into a step function
at $\beta=3^{3/4}$.

\subsection{The number of settings for a target coverage}\label{ss:cost}

Theorem~\ref{thm:threshold} gives the base $3^{3/4}$ at which the number of settings must grow with
$n$. It does not give the number of settings needed at a given $n$ and target coverage. Let
\begin{equation}\label{eq:kstar-def}
   k^\star(n,\delta):=\min\bigl\{k:\ \E[\Ncov(k)]\ge(1-\delta)4^n\bigr\}
\end{equation}
be the least number of uniformly random settings whose expected coverage reaches a fraction
$1-\delta$. The realized coverage is close to its expectation. Under sampling with replacement,
changing one setting changes $\Ncov$ by at most $2^n$, so the bounded-differences inequality
\cite{mcdiarmid1989bounded} gives
$\Prob\bigl[|\Ncov(k)-\E[\Ncov(k)]|\ge\tau4^n\bigr]\le2\exp(-2\tau^24^n/k)\le2\exp\bigl(-2\tau^2(4/3)^n\bigr)$
for every $\tau>0$ and $k\le3^n$. Hence $k^\star$ also describes a single draw of settings. Section~\ref{ss:opt} compares random sampling with other rules for choosing settings,
and there we write $k^\star_{\mathrm{rand}}$ for $k^\star$.

\begin{theorem}[Settings for a target coverage]\label{thm:kdelta}
Fix $\delta\in(0,1)$. Under uniform sampling of settings, with or without replacement, we have
\begin{equation}\label{eq:kstar-rand}
   \log_3 k^\star(n,\delta)=\frac{3n}{4}+\frac{\sqrt{3n}}{4}\,\zd
      -\frac{\zd^{2}-1}{12}-\frac{\gamma}{\ln 3}+\frac{c(\delta)}{\sqrt n}+\varepsilon_n(\delta),
\end{equation}
where
\begin{equation}\label{eq:cdelta}
   c(\delta):=\Bigl(\frac{\pi^2}{3\sqrt3\,\ln^23}+\frac{29\sqrt3}{432}\Bigr)\zd
      -\frac{11\sqrt3}{432}\,\zd^3\approx1.690\,\zd-0.0441\,\zd^3,
\end{equation}
and the remainder satisfies
\begin{equation}\label{eq:epslat}
   \limsup_{n\to\infty}|\varepsilon_n(\delta)|\ \le\ \epslat+3\times10^{-8}<2.4\times10^{-4},
   \qquad \epslat:=\frac{\bigl|\Gamma\bigl(1+2\pi i/\ln3\bigr)\bigr|}{\pi}.
\end{equation}
More precisely, $\varepsilon_n(\delta)=\lambda_n(\delta)+o(n^{-1/2})$, where the leading part of
$\lambda_n(\delta)$ is periodic in $\log_3k^\star$ with period one and amplitude $\epslat$.
\end{theorem}

\begin{proof}[Proof sketch]
We sketch the proof here, and Appendix~\ref{app:thm2} gives the details. By Lemma~\ref{lem:exp},
the uncovered fractions with and without replacement differ by a factor $1+o(n^{-1/2})$, because
$k3^{-n}$ decays exponentially in $n$. The density of $W$ at the crossover weight $\wstar=\log_3k^\star$ of \eqref{eq:wstar} is
$\Theta(n^{-1/2})$, so
this moves $\log_3k^\star$ by $o(1)$, and we solve $U(3^t)=\delta$ for $t=\log_3k$.

\emph{The step approximation.} Replacing the miss probability by the hard step
$\mathbf 1\{w>t\}$ changes $U(3^t)$ by $O(n^{-1/2})$ (Lemma~\ref{lem:step}), so
$\Prob[W>t]=\delta+O(n^{-1/2})$.
With quantile conversion (\stdref{std:quantile}) and the Berry--Esseen bound
(\stdref{std:berry}), this already provides $\log_3k^\star=\tfrac34n+\tfrac{\sqrt{3n}}4\zd+O(1)$.

\emph{The shape of the transition.} Near the crossover, $(1-3^{-w})^k\approx m(w-t)$ with the
\emph{miss profile} $m(s):=\exp(-3^{-s})$, which rises from $0$ to $1$ over a window of width
$O(1)$. The profile is the distribution function of a random variable $S$: a standard Gumbel
variable divided by $\ln3$, with mean $\mu_m=\gamma/\ln3$ and variance
$v_m=\pi^2/(6\ln^23)\approx1.363$. Replacing the sum over weights by an integral against the
Edgeworth density $f$ of $W$ (\stdref{std:berry}) and exchanging the order of integration gives
\[
   U(3^t)\approx\int_{-\infty}^{\infty}f(w)\,m(w-t)\,dw=\E\bigl[\bar F(t+S)\bigr],
   \qquad \bar F(x):=\int_x^\infty f(s)\,ds,
\]
with $S$ independent of $W$. Hence $U(3^t)$ is approximately $\Prob[W-S\ge t]$, and $\log_3k^\star$
is approximately the $(1-\delta)$-quantile of $W-S$. A second-order Taylor expansion of $\bar F$
about $t+\mu_m$ gives $\E[\bar F(t+S)]=\bar F(t+\mu_m)-\tfrac12v_mf'(t+\mu_m)+O(n^{-3/2})$. In
words, a smooth transition behaves like a hard step moved by the mean of the profile and widened by
its variance. Let $\widetilde Q_{1-\delta}$ be the $(1-\delta)$-quantile of the Edgeworth
approximation to $W$, that is, the point where $\bar F(\widetilde Q_{1-\delta})=\delta$. Near this
quantile $f'\approx-f\,\zd/\sigma_W$, and solving $U=\delta$ gives
$t=\widetilde Q_{1-\delta}-\gamma/\ln3+\zd v_m/(2\sigma_W)+o(n^{-1/2})$, up to the lattice
correction below. The Cornish--Fisher expansion (\stdref{std:cf}) gives
$\widetilde Q_{1-\delta}=\tfrac34n+\tfrac{\sqrt{3n}}4\zd-(\zd^{2}-1)/12
-\sqrt3(11\zd^3-29\zd)/(432\sqrt n)+O(1/n)$. With $\sigma_W=\sqrt{3n}/4$, the two terms of order
$n^{-1/2}$ add to $c(\delta)/\sqrt n$.

\emph{The lattice correction.} The weight takes integer values, so replacing the sum by an
integral leaves an error $E_n$. By Poisson summation (\stdref{std:poisson}), its leading part is a periodic
oscillation in $t$ of amplitude $f(t)\,\epslat$.  Solving for $t$ divides by $f(t)=\Theta(n^{-1/2})$, which turns this
into an oscillation of amplitude at most $\epslat$ in $\log_3k^\star$ that does not vanish as $n$
grows.
\end{proof}

In words, Theorem~\ref{thm:kdelta} says that $\log_3k^\star$ is, up to $\varepsilon_n(\delta)$,
the $(1-\delta)$-quantile of $W-S$. Here $W$ is the binomial weight of a uniformly drawn observable,
and $S$, independent of $W$, describes the gradual transition between covered and uncovered
weights. At $\delta=\tfrac12$, the $\sqrt n$ term and $c(\delta)$ vanish and the skewness term
equals $+\tfrac1{12}$. As $\delta\to0$, the skewness term grows in magnitude. The constant
$-\gamma/\ln3\approx-0.525$ is the mean of $-S$, and it lowers $k^\star$ by the factor $e^{-\gamma}$
relative to a hard step at the crossover weight. The variance $v_m\approx1.363$ of $S$ adds to the
variance $3n/16$ of $W$, and the wider distribution has its $(1-\delta)$-quantile further out, by
about $\zd v_m/(2\sigma_W)\approx1.574\,\zd/\sqrt n$. This is most of $c(\delta)/\sqrt n$. The rest,
$-\sqrt3(11\zd^3-29\zd)/(432\sqrt n)$, is the next Cornish--Fisher correction of the binomial weight
and comes from its kurtosis and squared skewness. The term $c(\delta)/\sqrt n$ raises $k^\star$ when
$\delta<\tfrac12$ and lowers it when $\delta>\tfrac12$. The cost of repeated hits is the factor
$e^{1-\gamma}$ against the counting bound in Section~\ref{ss:opt}. The constant $\gamma/\ln3$ and the amplitude
$\epslat$ also appear in the expected maximum of geometric random variables. If $G_1,\dots,G_k$ are
independent with $\Prob[G_j>w]=3^{-w}$, then $(1-3^{-w})^k=\Prob[\max_jG_j\le w]$, so
$U(k)=\Prob[\max_jG_j\le W]$ for $W$ independent of the $G_j$. As $k\to\infty$, the expected
maximum is $\log_3k+\gamma/\ln3+\tfrac12$ plus a periodic function of $\log_3k$ whose leading
Fourier component has amplitude $\epslat$, up to $o(1)$
\cite{szpankowski1990yet,eisenberg2008expectation,louchard2006asymptotics}. The
explicit terms cannot be replaced by the
$(1-\delta)$-quantile of $\Bin(n,\tfrac34)$ minus $\gamma/\ln3$: that quantile is an integer, while
$\log_3k^\star$ takes non-integer values that shift with $n$, so their difference does not
converge. For the same reason, maxima of independent geometric random variables such as the $G_j$ have no
limiting distribution other than a point mass, that is, a distribution concentrated on a single
value, under any centering and scaling
\cite{anderson1970extreme}.

Table~\ref{tab:residual} shows how much of the finite-$n$ error the term $c(\delta)/\sqrt n$
removes. We solved $U(k)=\delta$ numerically and subtracted the explicit terms of
\eqref{eq:kstar-rand} with and without $c(\delta)/\sqrt n$. At $\delta=0.1$ and $n=80$, the terms
without it underestimate $\log_3k^\star$ by $0.217$, a factor of $1.27$ in $k^\star$, and
$c(\delta)/\sqrt n=0.232$ accounts for all but $0.015$ of this. For $n\ge80$ the remaining error
$\varepsilon_n(\delta)$ is below $0.015$ in $\log_3k^\star$ at $\delta=0.1$, within $2\%$ in
$k^\star$, and below $0.09$ at $\delta=10^{-3}$, within $10\%$. It decays like $1/n$: the product
$n\,\varepsilon_n(\delta)$ lies between $-1.2$ and $-0.7$ at $\delta=0.1$ and between $-8.5$ and
$-7.1$ at $\delta=10^{-3}$. At $n\le20$ without replacement, the $k^\star$ settings are a sizable
fraction of all $3^n$, which lowers $k^\star$ by an amount comparable to $c(\delta)/\sqrt n$. At
$n=10$ and $\delta=0.1$, for example, \eqref{eq:kstar-rand} with $\varepsilon_n$ omitted
overestimates $k^\star$ by a factor of $1.43$. For sizing an experiment at such sizes,
\eqref{eq:Ecov} can be evaluated directly as a sum of $n+1$ terms, and
Theorem~\ref{thm:kdelta} describes how the result scales with $n$ and $\delta$.

\begin{table}[htbp]
\centering
\caption{The finite-$n$ error of Theorem~\ref{thm:kdelta} with and without the $n^{-1/2}$ term.
The first row of each block is the root $\log_3k$ of $U(k)=\delta$, for sampling with replacement,
minus the explicit terms of \eqref{eq:kstar-rand} other than $c(\delta)/\sqrt n$. The last row is
the remainder $\varepsilon_n(\delta)$, computed before rounding.}
\label{tab:residual}
\renewcommand{\arraystretch}{1.15}
\begin{tabular}{llccccc}
\toprule
 & & \multicolumn{5}{c}{$n$} \\
\cmidrule(l){3-7}
$\delta$ & & $80$ & $160$ & $320$ & $640$ & $1280$ \\
\midrule
$0.1$     & without $c(\delta)/\sqrt n$ & $0.217$  & $0.157$  & $0.113$  & $0.081$  & $0.057$  \\
          & $c(\delta)/\sqrt n$          & $0.232$  & $0.164$  & $0.116$  & $0.082$  & $0.058$  \\
          & $\varepsilon_n(\delta)$      & $-0.015$ & $-0.006$ & $-0.003$ & $-0.001$ & $-0.001$ \\
\midrule
$10^{-3}$ & without $c(\delta)/\sqrt n$ & $0.349$  & $0.262$  & $0.194$  & $0.142$  & $0.103$  \\
          & $c(\delta)/\sqrt n$          & $0.438$  & $0.310$  & $0.219$  & $0.155$  & $0.110$  \\
          & $\varepsilon_n(\delta)$      & $-0.089$ & $-0.048$ & $-0.026$ & $-0.013$ & $-0.007$ \\
\bottomrule
\end{tabular}
\end{table}

\subsection{Optimality of random sampling}\label{ss:opt}

A deliberate choice of settings could avoid the repeated hits of random sampling on observables
that are already covered. Such a choice cannot change the rate or the
$\sqrt n$ term, because the counting bound \eqref{eq:anybound}, which every collection of settings
obeys, already forces both.

\begin{theorem}[Converse]\label{thm:lb}
  The least number $k^\star_{\mathrm{opt}}(n,\delta)$ of settings for which some
collection covers a $1-\delta$ fraction of the observables satisfies
\begin{equation}\label{eq:kopt}
   \log_3 k^\star_{\mathrm{opt}}(n,\delta)=\frac{3n}4+\frac{\sqrt{3n}}4\zd+O(1).
\end{equation}
\end{theorem}

\begin{proof}[Proof sketch]
We sketch the proof here, and Appendix~\ref{app:thm3} gives the details. Dividing the counting
bound \eqref{eq:anybound} by $4^n$ shows that the covered fraction of any collection is at most
$\sum_wp_w\min(1,3^{\wstar-w})$. The
probabilistic method gives $k^\star_{\mathrm{opt}}\le k^\star$: at $k=k^\star$, the expected coverage
reaches $(1-\delta)4^n$, so some realization does. In the other direction, the part of the sum
above the crossover is at most $\tfrac32\pmax(n)=O(n^{-1/2})$ by \eqref{eq:fracpart}, so any
collection covering a $1-\delta$ fraction has $\Prob[W>\wstar]\le\delta+O(n^{-1/2})$, and the
quantile argument of Theorem~\ref{thm:kdelta} gives the same exponent.
\end{proof}

The bound \eqref{eq:anybound} charges nothing for overlap until a shell is fully covered, which is
why no selection rule can do better.

In finite-blocklength terms, $k^\star_{\mathrm{opt}}(n,\delta)$ is the least size of a code that
reproduces a uniformly drawn observable with zero distortion \eqref{eq:dontcare} except with
probability $\delta$. For a memoryless source at a distortion $d$ above the smallest one attainable
at finite rate, Kostina and Verd\'u \cite[Thm.~12]{kostina2012fixed}, and independently Ingber and
Kochman \cite{ingber2011dispersion}, show that the logarithm of the least code size is
$nR(d)+\sqrt{nV(d)}\,\Phi^{-1}(1-\delta)+O(\log n)$. The rate dispersion $V(d)$ is the variance of
the number of nats needed to reproduce one source symbol within distortion $d$. For covering, the
smallest distortion attainable at finite rate is zero, so their result does not apply. As $d\to0$, the number of nats for a
factor $P_i$ tends to $\ln3$ if $P_i\neq I$ and to $0$ if $P_i=I$, which gives
$V(0)=\tfrac3{16}\ln^23$. In base $3$, $nR(0)+\sqrt{nV(0)}\,\zd$ is
$\tfrac34n+\tfrac{\sqrt{3n}}4\zd$, and Theorem~\ref{thm:lb} shows that these two terms hold at zero
distortion with an $O(1)$ remainder.

Theorem~\ref{thm:lb} leaves an $O(1)$ gap in the exponent,
that is, a constant factor in $k$, and we can bound that factor explicitly. Let
\begin{equation}\label{eq:kcb-def}
   k^\star_{\mathrm{cb}}(n,\delta):=\min\Bigl\{k:\ \sum_{w=0}^{n}\binom nw\min(3^w,k)\ \ge\
   (1-\delta)4^n\Bigr\}
\end{equation}
be the least $k$ for which the counting bound \eqref{eq:anybound} permits a $1-\delta$ fraction, so
that $k^\star_{\mathrm{cb}}\le k^\star_{\mathrm{opt}}\le k^\star_{\mathrm{rand}}$.

\begin{corollary}[Random sampling against any selection rule]\label{cor:constant}
Fix $\delta\in(0,1)$. Then
\begin{equation}\label{eq:constant}
   \limsup_{n\to\infty}\log_3\frac{k^\star_{\mathrm{rand}}(n,\delta)}{k^\star_{\mathrm{cb}}(n,\delta)}
   \ \le\ 1-\frac{\gamma}{\ln3}+\epslat+3\times10^{-8} .
\end{equation}
In particular, $k^\star_{\mathrm{rand}}<3e^{-\gamma}\,3^{\epslat+3\times10^{-8}}\,k^\star_{\mathrm{opt}}
<1.69\,k^\star_{\mathrm{opt}}$ for all sufficiently large $n$.
\end{corollary}

\begin{proof}[Proof sketch]
We sketch the proof here, and Appendix~\ref{app:thm3} gives the details. At offset $s=w-\wstar$
from the crossover, random sampling covers a fraction $1-\exp(-3^{-s})$ of
the weight-$w$ shell, to leading order, while the counting bound allows a fraction
$\min(1,3^{-s})$. The corresponding miss profiles are $m(s)=\exp(-3^{-s})$ and
$m_{\mathrm{cb}}(s)=\max(0,1-3^{-s})$. The argument of Theorem~\ref{thm:kdelta} applies to both profiles, and
the crossover of each rule is the quantile $\widetilde Q_{1-\delta}$ from the proof of
Theorem~\ref{thm:kdelta}, minus the mean of its profile, up to a lattice correction.
The profile means are $\mu_m=\gamma/\ln3$ and
$\mu_{\mathrm{cb}}=\int_0^\infty s\ln3\cdot3^{-s}\,ds=1/\ln3$, so the crossovers differ by
$(1-\gamma)/\ln3$. The lattice correction for $m$ is at most $\epslat+3\times10^{-8}$, as in
Theorem~\ref{thm:kdelta}. For $m_{\mathrm{cb}}$, which has a kink at $s=0$, that is, a jump in its derivative, the lattice
correction tends to a periodic function of the crossover, which Appendix~\ref{app:thm3} finds in
closed form and which is never below $1/\ln3-1$. The counting-bound crossover is therefore at least
$\widetilde Q_{1-\delta}-1+o(1)$, while the random crossover is at most
$\widetilde Q_{1-\delta}-\gamma/\ln3+\epslat+3\times10^{-8}+o(1)$. Exponentiating the difference,
$k^\star_{\mathrm{rand}}/k^\star_{\mathrm{cb}}$ is eventually below
$3e^{-\gamma}\,3^{\epslat+3\times10^{-8}}\approx1.685$, and $k^\star_{\mathrm{cb}}\le
k^\star_{\mathrm{opt}}$ gives $k^\star_{\mathrm{rand}}<1.69\,k^\star_{\mathrm{opt}}$.
\end{proof}

\begin{figure}[!htb]
\centering
\includegraphics[width=0.6\linewidth]{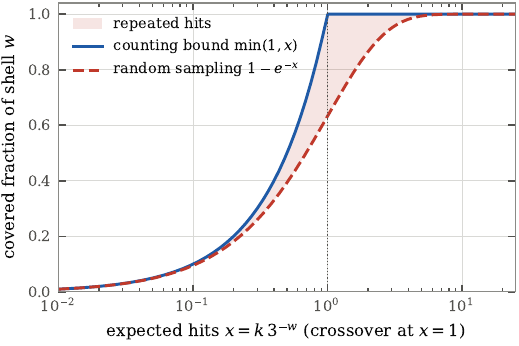}
\caption{Per-shell coverage against the expected number of hits $x=k3^{-w}$. The counting bound
\eqref{eq:anybound} reaches $\min(1,x)$, and random sampling reaches $1-e^{-x}$. The shaded area
between the curves is the coverage random sampling spends on repeated hits, and its mean offset
produces the factor $e^{1-\gamma}\approx1.53$ that underlies Corollary~\ref{cor:constant}.}
\label{fig:profile}
\end{figure}

The factor $e^{1-\gamma}\approx1.53$, the ratio of the two crossovers before their lattice
corrections, has a simple origin.
Figure~\ref{fig:profile} plots the per-shell coverage against the expected number of hits
$x=k3^{-w}$. The counting bound reaches $\min(1,x)$, while random sampling reaches only $1-e^{-x}$,
because some settings land on observables that are already covered. The same loss appears in the
coupon-collector problem, in which coupons are drawn at random until every one of $N$ types has
been seen. Covering the weight-$w$ observables on a fixed support of $w$ qubits by random settings
is this problem with $N=3^w$ coupons. The curve $1-e^{-x}$ is the
probability that a given coupon has been collected after draws that hit it $x$ times on average,
and $\gamma$ is the constant in the expected coupon-collection time $N\ln N+\gamma N+O(1)$, whose limit law Erd\H{o}s and R\'enyi derived
\cite{erdos1961classical}. The slope of $\min(1,x)$ jumps at $x=1$, and this kink makes
$k^\star_{\mathrm{cb}}$ oscillate with $n$ instead of converging,
and Corollary~\ref{cor:constant} bounds the ratio at the least favorable point of that
oscillation.

Table~\ref{tab:counts} compares $k^\star_{\mathrm{rand}}$ with $k^\star_{\mathrm{cb}}$ at $n=10$, $20$,
$50$ and $100$. At these sizes, the ratio lies between $1.44$ and $1.87$,
except at $n=10$ and $\delta=0.01$, where both rules need most of the $3^{10}$ settings and the
ratio is $1.02$. At $\delta=0.1$ the ratio oscillates with $n$, and at $\delta=0.01$ it increases over the sizes shown.
Several entries exceed $1.69$, because the constant of Corollary~\ref{cor:constant} holds in the
limit of large $n$, which the ratio approaches slowly. For a given $n$ and $\delta$, an experimenter
can compute $k^\star_{\mathrm{rand}}$ and $k^\star_{\mathrm{cb}}$ directly from \eqref{eq:Ecov} and
\eqref{eq:kcb-def}.

\begin{table}[!htb]
\centering
\caption{Settings needed to cover a $1-\delta$ fraction of the $4^n$ observables. The
$k^\star_{\mathrm{rand}}$ columns give $k^\star$ of \eqref{eq:kstar-def} without replacement,
computed from \eqref{eq:Ecov}. Among the tabulated sizes, the value with replacement differs by more than $1\%$ only at
$n=10$ and $n=20$. The $k^\star_{\mathrm{cb}}$ columns give the counting bound \eqref{eq:kcb-def}, a lower bound
on the number of settings any selection rule needs, and the ratio is
$k^\star_{\mathrm{rand}}/k^\star_{\mathrm{cb}}$.}
\label{tab:counts}
\renewcommand{\arraystretch}{1.15}
\begin{tabular}{r@{\qquad}ccc@{\qquad}ccc}
\toprule
 & \multicolumn{3}{c}{$\delta=0.1$} & \multicolumn{3}{c}{$\delta=0.01$} \\
\cmidrule(r{1.5em}){2-4}\cmidrule{5-7}
$n$ & $k^\star_{\mathrm{cb}}$ & $k^\star_{\mathrm{rand}}$ & ratio
    & $k^\star_{\mathrm{cb}}$ & $k^\star_{\mathrm{rand}}$ & ratio \\
\midrule
$10$  & $1.4\times10^{4}$  & $2.0\times10^{4}$  & $1.44$ & $4.9\times10^{4}$  & $4.9\times10^{4}$  & $1.02$ \\
$20$  & $1.0\times10^{8}$  & $1.7\times10^{8}$  & $1.71$ & $7.5\times10^{8}$  & $1.2\times10^{9}$  & $1.56$ \\
$50$  & $2.6\times10^{19}$ & $4.1\times10^{19}$ & $1.61$ & $6.5\times10^{20}$ & $1.1\times10^{21}$ & $1.73$ \\
$100$ & $1.1\times10^{38}$ & $1.8\times10^{38}$ & $1.60$ & $1.1\times10^{40}$ & $2.0\times10^{40}$ & $1.87$ \\
\bottomrule
\end{tabular}
\end{table}

\subsection{Observables of bounded weight}\label{ss:bounded}

Many quantities of experimental interest involve only Pauli observables of bounded weight, and for
these observables the rate in Theorem~\ref{thm:threshold} changes.
For example, a local Hamiltonian is a sum of Pauli terms on a few neighboring sites.
Two-point correlators, local order parameters and stabilizer generators are also of low weight.
Classical shadows are analyzed with this bounded-weight
restriction in mind \cite{huang2020predicting}.

\begin{corollary}[Bounded weight]\label{cor:bounded}
Fix $\omega\in(0,\tfrac34)$ and restrict attention to observables of weight at most $\omega n$. Let
$k^\star_\omega(n,\delta)$ be the least $k$ for which the expected covered fraction of these
observables under uniform sampling is at least $1-\delta$.
\begin{enumerate}
\item[\textup{(i)}] The critical growth base is $3^{\omega}{:}$ for $k=\beta^{n}$, the covered
fraction of these observables tends to $1$ if $\beta>3^{\omega}$ and to $0$ if
$\beta<3^{\omega}$.
\item[\textup{(ii)}] For fixed $\delta\in(0,1)$, we have
\begin{equation}\label{eq:kstar-bounded}
   \log_3 k^\star_\omega(n,\delta)=\omega n+O(1).
\end{equation}
If instead $\delta=\delta_n\to0$ with $\ln(1/\delta_n)=3^{o(n)}$, then
$\log_3 k^\star_\omega(n,\delta_n)=\omega n+\log_3\ln(1/\delta_n)+O(1)$, where the $O(1)$ term depends on
$\omega$ only.
\end{enumerate}
\end{corollary}

The rate drops to $3^\omega$ because the shell sizes $\binom nw3^w$ increase up to weight
$\tfrac34n$. Among observables of weight at most $\omega n$, most therefore have weight close to the top
weight $\omega n$, and these play the role that the shell of weight $\tfrac34n$ plays in
Theorem~\ref{thm:threshold}.
The $\sqrt n$ term disappears because the ratio of consecutive shell sizes,
$p_{w+1}/p_w=3(n-w)/(w+1)$, stays above a constant larger than one near $w=\omega n$. The weight of
a random low-weight observable therefore falls off geometrically below the top weight and has a
spread of $O(1)$ rather than $\Theta(\sqrt n)$, so changing the coverage target moves the crossover
by only $O(1)$.
The proof is in Appendix~\ref{app:cor}.

\section{The design effect of a measurement setting}\label{sec:design}

In this section, we study what a setting provides to an estimator. The $2^n$ estimates of Pauli
expectations from one setting all come from one histogram of outcomes, so they are correlated, and
we express that correlation through a single number, the design effect of the setting. We compute
the mean design effect over settings for any state, evaluate this mean for Haar-random, GHZ and W
states, and show how to estimate the design effect of a setting from the counts without knowing the
state.

\subsection{The covariance of the Pauli estimates from one setting}\label{ss:spectrum}

Fix a setting $M$ and index the $2^n$ observables it covers by their support $S\subseteq[n]$, so
that an observable $P_S$ agrees with $M$ on $S$ and is the identity elsewhere. These observables commute and are
diagonal in the product eigenbasis of $M$,
\begin{equation}\label{eq:eig}
   P_S\,|b\rangle=(-1)^{|S\cap b|}\,|b\rangle,\qquad b\in\{0,1\}^n,
\end{equation}
where $|S\cap b|$ counts the coordinates in $S$ at which $b$ equals $1$. Let $\Wal$ be the
$2^n\times2^n$ Walsh--Hadamard matrix \cite{odonnell2014analysis} with entries $\Wal_{S,b}=(-1)^{|S\cap b|}$, which is symmetric and
satisfies $\Wal^2=2^nI$.

Let $p$ be the outcome distribution of the setting $M$ on the target state and $\phat$
the empirical frequencies from $\Nsh$ shots, so that $\Nsh\phat\sim\mathrm{Multinomial}(\Nsh,p)$.
By \eqref{eq:eig}, the vector of true expectations is $g=\Wal p$ and the vector of estimates is
$\hat y=\Wal\phat$, that is,
\begin{equation}\label{eq:wht}
   \langle P_S\rangle=\sum_b(-1)^{|S\cap b|}p(b),\qquad
   p(b)=2^{-n}\sum_S(-1)^{|S\cap b|}\langle P_S\rangle.
\end{equation}
Since $\operatorname{Cov}(\phat)=(\operatorname{diag}p-pp^\top)/\Nsh$, the covariance of the
estimates is
\begin{equation}\label{eq:C-def}
   C:=\operatorname{Cov}(\hat y)=\frac1{\Nsh}\Bigl(\Wal\operatorname{diag}(p)\Wal-gg^\top\Bigr),
\end{equation}
and reading off an entry gives
\begin{equation}\label{eq:cov-collision}
   \operatorname{Cov}(\hat y_S,\hat y_T)
      =\frac1{\Nsh}\Bigl(\langle P_{S\triangle T}\rangle-\langle P_S\rangle\langle P_T\rangle\Bigr),
\end{equation}
where $S\triangle T$ is the symmetric difference. The off-diagonal entries are expectations of the
target state itself, so the correlation is a property of the state and of the setting's basis, not
of the noise. The identity \eqref{eq:cov-collision} is the one underlying the covariance-aware
grouping methods of Section~\ref{ss:prior}.

The form \eqref{eq:C-def} makes the spectrum easy to read off. Writing $h_b$ for the column of
$\Wal$ indexed by $b$, we have $\Wal\operatorname{diag}(p)\Wal=\sum_bp(b)\,h_bh_b^\top$, and the
columns are orthogonal with $h_b^\top h_b=2^n$. Apart from the rank-one term $gg^\top$, the
eigenvalues of $\Nsh C$ are the outcome probabilities $2^np(b)$. A standard measure of how
many independent noise components a covariance matrix has is its participation ratio
$\mathrm{PR}(C)=(\operatorname{tr}C)^2/\operatorname{tr}[C^2]$, which equals $m$ when $C$ has $m$
equal nonzero eigenvalues. Using $\|g\|^2=2^n\sum_bp(b)^2$ and $g^\top\Wal\operatorname{diag}(p)\Wal
g=4^n\sum_bp(b)^3$, the participation ratio is
\begin{equation}\label{eq:PRexact}
   \mathrm{PR}(C)=\frac{\bigl(1-\sum_bp(b)^2\bigr)^2}
               {\sum_bp(b)^2-2\sum_bp(b)^3+\bigl(\sum_bp(b)^2\bigr)^2}.
\end{equation}
The participation ratio is close to the inverse collision probability of the outcome
distribution in the cases we computed, and the inverse collision probability has simpler closed
forms. We therefore work with the \emph{effective support size} of the outcome distribution of the
setting,
\begin{equation}\label{eq:nu-collision}
   \nueff:=\Bigl(\sum_bp(b)^2\Bigr)^{-1}=2^{H_2(p)},\qquad H_2(p):=-\log_2\sum_bp(b)^2 .
\end{equation}
Here $H_2$ is the R\'enyi entropy of order two \cite{renyi1961measures}, also called the collision
entropy \cite{bosyk2012collision}, because $\sum_bp(b)^2$ is the probability that two independent
shots return the same outcome. Since $\sum_bp(b)^3\ge(\sum_bp(b)^2)^2$, we have
$\mathrm{PR}(C)\ge\nueff-1$ for every $p$ supported on more than one outcome, with equality when $p$ is uniform
on its support, where the missing direction is the noiseless identity observable. We evaluated \eqref{eq:PRexact} and \eqref{eq:nu-collision} for all $3^n$ settings of the GHZ and W
states and of one Haar-random draw at $n=4,5,6$. The difference $|\mathrm{PR}(C)-\nueff|$ is at most $1.94$
over all of these settings, while $\nueff$ itself ranges up to $2^n$. This closeness is a
numerical observation and does not hold for every $p$: one outcome of probability $q$ together with
a flat tail over many outcomes gives $\mathrm{PR}(C)\to(1+q)^2/q^2$, against $\nueff\to1/q^2$.

In the cases we computed, the $2^n$ estimates from a setting are therefore driven by approximately
$\nueff$ independent noise components. We call
\begin{equation}\label{eq:D-def}
   D:=\frac{2^n}{\nueff}=2^n\sum_bp(b)^2
\end{equation}
the \emph{design effect} of the setting. The name comes from survey sampling
\cite{kish1965survey}, where the design effect measures the loss of efficiency of a clustered sample
relative to a simple random sample of the same size. Here a setting plays the role of the cluster. We borrow the formula and the name: our $D$ is not
the variance-inflation factor of a particular estimator.
With the $2^n$ outcome probabilities $p(b)$ in the role of weights, \eqref{eq:D-def} is Kish's
design effect for unequal weighting, $m\sum_iw_i^2/(\sum_iw_i)^2$ for $m$ sample units with
weights $w_i$ \cite{kish1992weighting}, and $\nueff$ is the effective sample size of importance
sampling \cite{kong1994sequential}.
Apart from the rank-one term $gg^\top$, the eigenvalues $2^np(b)$ of $\Nsh C$ have mean one, so
$D$ is their mean square and $D-1$ is their variance. A design effect of one means white noise, for
which a least-squares fit loses nothing by ignoring the correlation, and $D-1$ measures how far a
setting is from this case.
For a design that weights all settings equally, the relevant aggregate is the mean over uniformly
random settings,
\begin{equation}\label{eq:Dbar}
   \bar D:=\E[D]=2^n\,\E\bigl[\nueff^{-1}\bigr].
\end{equation}

The design effect can be written through the expectations of the covered observables, which
gives the mean design effect over settings in closed form for any state.
\begin{lemma}[Design effect and Pauli expectations]\label{lem:sector}
Every setting satisfies $D=\sum_{S\subseteq[n]}\langle P_S\rangle^2$, where $P_S$ runs over the
$2^n$ observables covered by the setting, including $P_\emptyset=I^{\otimes n}$. For every $n$-qubit
state $\rho$, the mean design effect \eqref{eq:Dbar} over uniformly random settings is
\begin{equation}\label{eq:Dbar-sector}
   \bar D=\sum_{P\in\Psf_n}3^{-|P|}\langle P\rangle^2=\sum_{w=0}^{n}3^{-w}A_w
         =\Bigl(\frac23\Bigr)^{n}\sum_{S\subseteq[n]}\operatorname{tr}\rho_S^2,
\end{equation}
where $A_w:=\sum_{|P|=w}\langle P\rangle^2$ is the sector length of weight $w$
\cite{wyderka2020characterizing}, $\rho_S$ is the reduced state of the qubits in $S$, and
$\operatorname{tr}\rho_\emptyset^2=1$.
\end{lemma}
\begin{proof}
Since $g=\Wal p$ and $\Wal^2=2^nI$, Parseval's identity for the Walsh--Hadamard transform
\cite{odonnell2014analysis} gives $\sum_S\langle P_S\rangle^2=\|g\|^2=2^n\sum_bp(b)^2=D$. A uniformly
random setting covers $P$ with probability $3^{-|P|}$ by \eqref{eq:F1}, which gives the first form
of \eqref{eq:Dbar-sector}, and grouping by weight gives the second. For the third, the Pauli
expansion of $\rho_S$ gives
$\operatorname{tr}\rho_S^2=2^{-|S|}\sum\langle P\rangle^2$, summed over the observables supported
in $S$. Summing over $S$ counts an observable of weight $w$ with total weight
$\sum_{S\supseteq\operatorname{supp}P}2^{-|S|}=2^{-w}(3/2)^{n-w}=(3/2)^n3^{-w}$. Hence
$(2/3)^n\sum_S\operatorname{tr}\rho_S^2$ equals the first form.
\end{proof}
\begin{proposition}[What the design effect measures]\label{prop:mass}
Let $\rho$ be an $n$-qubit state other than the maximally mixed state, and let $D$ be the design
effect of a setting $M$.
\begin{enumerate}
\item[\textup{(i)}] $D-1=(2^n\operatorname{tr}\rho^2-1)\,C_\rho(M)$, where $C_\rho(M)$ is the
fraction of $\sum_{P\neq I^{\otimes n}}\langle P\rangle^2$ carried by the observables that $M$ covers.
\item[\textup{(ii)}] The total variance of the $2^n$ estimates is $\operatorname{tr}C=(2^n-D)/\Nsh$.
\item[\textup{(iii)}] For $\Nsh\ge2$, the estimate $\hat D$ obtained by multiplying
\eqref{eq:collision-est} by $2^n$ satisfies
$\hat D=\sum_{S}(\Nsh\hat y_S^2-1)/(\Nsh-1)$, and each summand is an unbiased estimate of
$\langle P_S\rangle^2$.
\end{enumerate}
\end{proposition}
\begin{proof}
(i) By Lemma~\ref{lem:sector}, $D=1+\sum_{S\neq\emptyset}\langle P_S\rangle^2$, and the Pauli
expansion of $\rho$ gives $\sum_{P\in\Psf_n}\langle P\rangle^2=2^n\operatorname{tr}\rho^2$. (ii) The
estimate $\hat y_S$ is the mean of $\Nsh$ independent $\pm1$ outcomes with expectation
$\langle P_S\rangle$, so its variance is $(1-\langle P_S\rangle^2)/\Nsh$, and summing over $S$ gives
$(2^n-D)/\Nsh$. (iii) Parseval's identity gives
$\sum_S\hat y_S^2=2^n\sum_b\phat(b)^2=2^n\sum_bc_b^2/\Nsh^2$, where $c_b$ is the number of shots with
outcome $b$, and substituting this into the sum gives $2^n\sum_bc_b(c_b-1)/(\Nsh(\Nsh-1))$.
Unbiasedness follows from $\E[\hat y_S^2]=\langle P_S\rangle^2+(1-\langle P_S\rangle^2)/\Nsh$.
\end{proof}
A large design effect therefore does not mean noisy estimates. By part~(i), $D$ is the squared
Pauli expectation that the setting captures, and by part~(ii) the total variance of its estimates
decreases as $D$ grows. What a large $D$ indicates is that the outcome distribution is concentrated,
so that the $2^n$ estimates are redundant. For the GHZ state of Proposition~\ref{prop:design}
measured in the all-$Z$ setting, $\nueff=2$ and $D=2^{n-1}$: the even-weight $Z$ strings have
expectation one and zero variance, and the odd-weight $Z$ strings all equal the same random sign in
every shot.

States with many large low-weight expectations have a large mean design effect. For a
pure product state, every reduced state $\rho_S$ is pure, so $\operatorname{tr}\rho_S^2=1$ for
every $S$ and $\bar D=(4/3)^n$. The eigenbases of the Pauli matrices $X$, $Y$ and
$Z$ form a 2-design on each qubit, that is, they reproduce the second moments of a Haar-random basis
\cite{elben2023toolbox}. The mean $\bar D$ is therefore $2^n$ times the collision probability
averaged over local Haar-random bases, and the last form of \eqref{eq:Dbar-sector} also follows
from the second-moment identities of randomized measurements
\cite{brydges2019probing,elben2019statistical,ketterer2019characterizing}.

\subsection{Haar-random, GHZ and W states}\label{ss:targets}

Averaged over Haar-random states, the design effect is less than $2$. For the GHZ and W states, the
mean design effect over settings grows like $\tfrac12(4/3)^n$.

\begin{proposition}[Design effect of Haar-random, GHZ and W states]\label{prop:design}
Let $D$ be the design effect \eqref{eq:D-def} of a setting, and $\nueff$ the effective support size
\eqref{eq:nu-collision} of its outcome distribution.
\begin{enumerate}
\item[\textup{(i)}] If the target is a Haar-random pure state on $n$ qubits, then for every setting
the expected design effect over the Haar measure is $\E[D]=2\cdot2^n/(2^n+1)$.
\item[\textup{(ii)}] If the target is the GHZ state $(|0\rangle^{\otimes n}+|1\rangle^{\otimes
n})/\sqrt2$ and the setting has $j\ge1$ factors equal to $Z$, then $\nueff=2^{n+1-j}$. If $j=0$,
then $\nueff=2^{n-1}$ when the number of $Y$ factors is even and $\nueff=2^n$ when it is odd. The mean
design effect over uniformly random settings is
\begin{equation}\label{eq:ghzD}
   \bar D_{\mathrm{GHZ}}=\frac12\Bigl(\frac43\Bigr)^{n}+\Bigl(\frac23\Bigr)^{n}.
\end{equation}
\item[\textup{(iii)}] If the target is the W state $n^{-1/2}\sum_{i=1}^{n}|e_i\rangle$, where
$e_i\in\{0,1\}^n$ has a single $1$ in position $i$, then the mean design effect over uniformly
random settings is
\begin{equation}\label{eq:wD}
   \bar D_{\mathrm W}=\frac12\Bigl(\frac43\Bigr)^{n}\Bigl(1+\frac1n\Bigr).
\end{equation}
\end{enumerate}
\end{proposition}

\begin{proof}
(i) For a Haar-random state on dimension $d=2^n$, the outcome probabilities in any fixed basis are
uniformly distributed on the probability simplex, whose large-$d$ limit is the Porter--Thomas law
\cite{porter1956fluctuations,boixo2018characterizing}. Hence $\E[p(b)^2]=2/(d(d+1))$, so
$\E\sum_bp(b)^2=2/(d+1)$ and $\E[D]=2^n\cdot2/(2^n+1)$ by \eqref{eq:D-def}.

(ii) Measuring $Z$ on $j\ge1$ qubits of the GHZ state gives either all zeros or all ones on those
qubits and leaves the other $n-j$ qubits in a product state, which the $X$ and $Y$ factors measure
with uniform outcomes. The outcome distribution is therefore uniform on $2\cdot2^{n-j}$ strings, so
$\nueff=2^{n+1-j}$. With $j=0$, the outcome is uniform on the $2^{n-1}$ strings of one parity when
the number of $Y$ factors is even, and uniform on all $2^n$ strings when it is odd.  For the mean, every nonempty proper subset $S$
has $\operatorname{tr}\rho_S^2=\tfrac12$, so $\sum_S\operatorname{tr}\rho_S^2=2^{n-1}+1$, and
Lemma~\ref{lem:sector} gives \eqref{eq:ghzD}.

(iii) For a subset $S$ of $a$ qubits with $0<a<n$, the reduced state of the W state is the
mixture $\tfrac an|\mathrm W_a\rangle\langle\mathrm W_a|+(1-\tfrac an)|0\rangle\langle0|^{\otimes a}$
of two orthogonal pure states, where $\mathrm W_a$ is the W state on $S$. Hence
$\operatorname{tr}\rho_S^2=(a^2+(n-a)^2)/n^2$, which also holds for $a=0$ and $a=n$. Summing over
subsets with $\sum_a\binom na a^2=n(n+1)2^{n-2}$ gives
$\sum_S\operatorname{tr}\rho_S^2=2^{n-1}(1+1/n)$, and Lemma~\ref{lem:sector} gives \eqref{eq:wD}.
\end{proof}

\begin{figure}[!htb]
\centering
\includegraphics[width=0.6\linewidth]{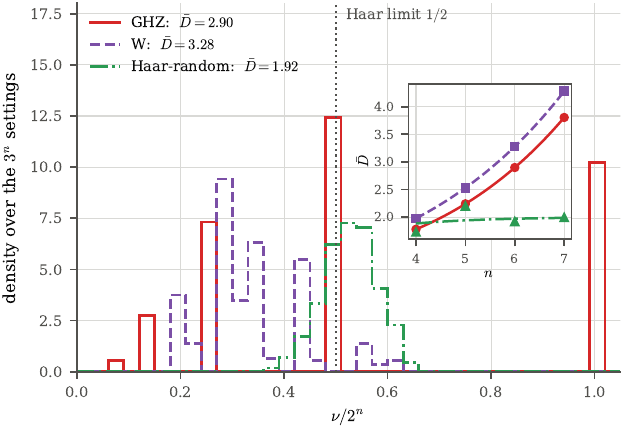}
\caption{Distribution of the effective support size $\nueff$ of
\eqref{eq:nu-collision} over all $3^n$ settings at $n=6$, as a fraction of the $2^n$ observables a
setting covers. The Haar-random state concentrates near $\tfrac12$, while the GHZ state spreads
over $[2^{1-n},1]$ according to Proposition~\ref{prop:design}. Inset: the mean design effect $\bar
D$ of \eqref{eq:Dbar} against $n$. Markers are
exhaustive enumerations over all $3^n$ settings, with one Haar-random draw at each $n$, and the
lines are \eqref{eq:ghzD}, \eqref{eq:wD} and the Haar average $2\cdot2^n/(2^n+1)$ of
Proposition~\ref{prop:design}(i). The covering analysis
gives all of these settings the same value, and the state does not.}
\label{fig:collision}
\end{figure}

Averaged over Haar-random states, the design effect is $1.882$, $1.939$, $1.969$ and $1.984$ at
$n=4,5,6,7$. These values increase toward the limit $2$ without reaching it, so the Haar-averaged design effect
stays bounded as $n$ grows: reading $2^n$ commuting
observables from one histogram of an unstructured state costs a factor of two. The corresponding
collision probability is $2/(2^n+1)$. Random-circuit studies use this Porter--Thomas value as the
benchmark for anticoncentration, that is, for outcome probabilities that are spread over all
bit-strings \cite{hangleiter2018anticoncentration,dalzell2022random}. The value $2\cdot2^n/(2^n+1)$ is an
average over the Haar measure. The mean over settings of a single Haar-random state fluctuates
around this average and can exceed $2$: enumerating all $3^n$ settings of one draw gave $\bar D=1.729$,
$2.197$, $1.917$ and $1.993$ at $n=4,\dots,7$. Over $100$ draws, the standard deviation of $\bar D$
is $0.13$, $0.09$, $0.06$ and $0.04$, so a single draw is close to the average for $n\ge6$.  For the GHZ state, \eqref{eq:ghzD} gives $1.778$, $2.239$, $2.897$ and $3.804$ at
$n=4,5,6,7$, which exhaustive enumeration reproduces, and $\bar D$ grows like $\tfrac12(4/3)^n$.
Here $\nueff$ spreads over the whole range from $2$ to $2^n$.    For the W state, \eqref{eq:wD} gives
$\bar D=1.975$, $2.528$, $3.278$ and $4.281$ at $n=4,5,6,7$, which exhaustive enumeration
reproduces, and $\bar D_{\mathrm W}$ grows like $\tfrac12(4/3)^n$, as the GHZ value does. Figure~\ref{fig:collision} shows how $\nueff$ is
distributed across settings at $n=6$.

\subsection{Estimating the design effect from data}\label{ss:online}

The values above assume a known target, but in tomography the target is what we are trying to
learn. Estimating the design effect does not require knowledge of the target state. By \eqref{eq:D-def}, $D$ depends on the outcome
distribution only through the collision probability $\sum_bp(b)^2$, which can be estimated from the
histogram the reconstruction already uses. If $c_b$ of the $\Nsh$ shots land on outcome $b$, the
fraction of coinciding pairs of shots,
\begin{equation}\label{eq:collision-est}
   \widehat{\Bigl(\sum_bp(b)^2\Bigr)}:=\frac{\sum_b c_b(c_b-1)}{\Nsh(\Nsh-1)},
\end{equation}
has expectation $\sum_bp(b)^2$ for every $\Nsh\ge2$ and every $p$, because two independent shots
coincide with that probability.   This is Simpson's unbiased estimator of his diversity index
\cite{simpson1949measurement}, now the standard unbiased estimator of a collision probability
\cite{acharya2017estimating}. The randomized-measurement toolbox uses the same statistic to
estimate purities and second R\'enyi entropies \cite{elben2019statistical,brydges2019probing}. Here
we apply it to a single Pauli setting. Multiplying by $2^n$ gives an unbiased estimate $\hat D$ of
the design effect of each setting, computed in one pass over the counts with no extra circuit runs.
Tiffeau-Mayer \cite{tiffeaumayer2024unbiased} gives an unbiased estimator of the variance of
\eqref{eq:collision-est}, which would supply error bars for $\hat D$.
The reciprocal $\hat\nueff=2^n/\hat D$ is not unbiased, and the plug-in $\sum_b\phat(b)^2$ is biased
upward by $(1-\sum_bp(b)^2)/\Nsh$. For an outcome distribution that is spread
over all $2^n$ strings, two shots rarely coincide unless $\Nsh$ is of order $2^{n/2}$, so $\hat D$ of
a single setting is informative only above that shot count. In the six-qubit simulations of
Section~\ref{sec:numerics}, the relative root-mean-square error of $\hat D$ for one setting is
below $10\%$ at $\Nsh=512$ and below $3\%$ at $\Nsh=8192$. The reconstructions there do not use $\hat D$. We use
the mean design effect to read how much an estimator gains by modeling the correlation.

\section{Tomography simulations}\label{sec:numerics}

Simulated six-qubit tomography shows what the correlation within a setting costs a
reconstruction, and what measuring settings saves in circuit runs. We compare a design that
measures observables one at a time with a design that measures settings, first at a matched number
of observables and then at a matched number of circuit runs.

\subsection{Designs, estimators and solvers}\label{ss:matched}

We compare two ways of spending a measurement budget on the same target state, with $\Nsh$ shots
per measurement. The \emph{observable design} draws $\Mobs$ distinct Pauli observables uniformly
from $\Psf_n\setminus\{I^{\otimes n}\}$ and measures each separately, so its errors are independent.
This is the design that compressed-sensing and message-passing analyses assume
\cite{gross2010compressed,siekierski2025approximate}. The \emph{setting design} draws $k$ distinct settings
uniformly and obtains the expectations of all $2^n-1$ covered non-identity observables by
marginalizing each histogram. This is the design of the experiments of Riofr\'io et al.\
\cite{riofrio2017experimental} and Steffens et al.\ \cite{steffens2017experimentally}. We fix $k$
and set $\Mobs$ to the expected number of non-identity observables that the $k$ settings cover,
$\E[\Ncov(k)]-1$ from \eqref{eq:Ecov}, rounded to an integer. We simulate each
setting by inverting \eqref{eq:wht} to obtain $p$ with no sampling error, drawing counts from
$\mathrm{Multinomial}(\Nsh,p)$, and transforming back, so that observables sharing a setting carry
the correlation \eqref{eq:cov-collision}.

At matched observable count, the observable and setting designs use $\Mobs\Nsh$ and $k\Nsh$ circuit runs, so the
setting design is cheaper by a factor $\Mobs/k$. In return, each measured observable is estimated
several times. Each of the $k$ settings covers $2^n-1$ non-identity observables, so the mean number
of estimates per measured observable is
\begin{equation}\label{eq:multiplicity}
   \cbar=\frac{k\,(2^n-1)}{\Ncov(k)-1},
\end{equation}
and with $\Mobs=\Ncov(k)-1$,
\begin{equation}\label{eq:tradeoff}
   \frac{\Mobs}{k}\cdot\cbar=2^n-1 .
\end{equation}
The saving in runs and the multiplicity trade off within the $2^n$ observables covered by a setting.
Equation \eqref{eq:tradeoff} is an identity, so the factor $\Mobs/k$ is bookkeeping relative to the
independent-observable model. Tomography experiments already measure settings and keep the full
histograms \cite{haffner2005scalable,riofrio2017experimental}. The simulations test how a
reconstruction should use those histograms.  At the scale of Theorem~\ref{thm:kdelta}, with
$k=3^{3n/4+O(\sqrt n)}$ and $\Mobs=(1-\delta)4^n$, both factors are exponential:
$\Mobs/k=1.75^{n+o(n)}$ and $\cbar=1.14^{n+o(n)}$.

We run three estimators on the setting design. Write $c_P$ for the number of the $k$ settings
that cover an observable $P$, and $\bar y_P$ for the average of its $c_P$ estimates.
\emph{Pooling} fits the averages $\bar y_P$ with equal weights. The average of $c_P$ estimates
has about $1/c_P$ of the variance of one estimate, and pooling discards this. \emph{Stacking}
fits the outcome histograms directly, by least squares on
$\sum_{s,b}(\phat_s(b)-p_s(b))^2$, where $s$ runs over the $k$ settings and $p_s$ is the outcome
distribution of setting $s$ under the candidate state. By \eqref{eq:wht} and Parseval's identity,
this loss equals $2^{-n}\sum_Pc_P(\bar y_P-\langle P\rangle)^2$ plus a term that does not depend on
the state. Stacking is therefore pooling with each observable weighted by its multiplicity, and like
pooling it treats the estimates from one setting as independent.

\emph{Biwhitening} also fits the histograms, after rescaling the count of outcome $b$ in setting
$s$ by a separable factor $u_sv_b$ chosen so that the noise variance has unit average in every row and
column \cite{landa2022biwhitening}. The factors are estimated from the counts through the binomial
relation between mean and variance, so no knowledge of the state is needed. The multinomial noise
of one histogram has variance close to $p_s(b)/\Nsh$ in each outcome, so weighting each outcome by
$p_s(b)^{-1/2}$ whitens the covariance \eqref{eq:C-def} up to its rank-one term. A separable factor
reproduces this weight only when $p_s(b)$ is the product of a function of $s$ and a function of
$b$. Biwhitening therefore removes the correlation for a state whose outcome distribution is the
same in every setting, and otherwise acts mainly through the multiplicity weighting it shares with
stacking. Because the rescaling attaches one factor to each
setting and one to each outcome, the rows of the rescaled operator stay on a common scale.
Approximate message passing assumes rows on a common scale in its Onsager correction, the memory
term that keeps the effective noise at each iteration close to Gaussian
\cite{donoho2009message,siekierski2025approximate}. We also tried whitening each setting fully with
the covariance \eqref{eq:C-def}, which is singular because the
identity observable carries no noise. Full whitening makes the noise isotropic, but it spreads the
row norms of the operator by an order of magnitude, and approximate message passing then failed in $71\%$ to $100\%$ of the $n=6$ runs,
against none with biwhitening.

Two references isolate the correlation. The \emph{maximum-likelihood} (ML) estimate maximizes the
multinomial likelihood of the histograms. It weights each outcome by the inverse of its model
probability and so accounts for both the multiplicity and the correlation, but it needs an exact
measurement model (Section~\ref{ss:readout}), and we use it only as a reference. The
\emph{independent design} measures each covered observable $P$ in $c_P$ separate experiments of
$\Nsh$ shots, so it has the multiplicity of the setting design without its correlation. It costs
about $2^n-1$ times more runs, and we fit it by least squares.

We reconstruct with Riemannian gradient descent (RGD) on the rank-one manifold
\cite{hsu2024quantum}, and with the approximate message passing (AMP) solver of Siekierski et al.\
\cite{siekierski2025approximate}. At iteration $t$, AMP soft-thresholds the eigenvalues of its
iterate at $\alpha\sigma_t\sqrt{2^n}$, where $\sigma_t$ is the root-mean-square residual. The
solver's default is $\alpha=2$, and at $n=6$ we use $\alpha=3$ for every design and estimator
(Section~\ref{ss:matched-results}). The targets are the GHZ, W and Haar-random states, and the
product state whose qubits all have Bloch vector $(1,1,1)/\sqrt3$. This product state has the same
outcome distribution in every setting, and the largest mean design effect of the four,
$\bar D=(4/3)^6=5.62$ (Section~\ref{ss:spectrum}). The simulations use $n=6$, coverage fractions
$\Mobs/4^n\in\{0.25,0.50,0.75,0.93\}$, reached with $k=27$, $86$, $226$ and $500$ settings, shots
$\Nsh\in\{512,2048,8192\}$, and $50$ trials per configuration. We report medians and quartiles
rather than means. At $25\%$ coverage the error on the GHZ state is bimodal for every estimator.
The GHZ state and $(|0\rangle^{\otimes n}-|1\rangle^{\otimes n})/\sqrt2$ differ only in the
$2^{n-1}$ observables of weight $n$ built from $X$ and $Y$ with an even number of $Y$, and $27$
random settings contain none of them with probability $(1-2^{n-1}/3^n)^{27}\approx0.30$. A mean
describes neither case.

\begin{figure}[!tb]
\centering
\includegraphics[width=\linewidth]{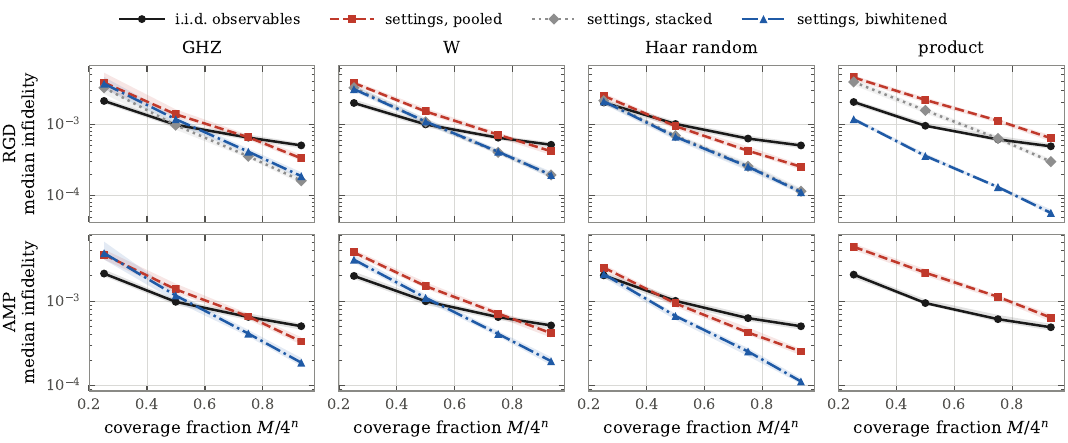}
\caption{Median infidelity against coverage fraction at $n=6$ and $\Nsh=2048$ shots per measurement, with the
interquartile range shaded, over $50$ trials. The observable and setting designs measure the same number $\Mobs$ of
distinct observables with the same shots per measurement. Top row RGD, bottom row AMP with
$\alpha=3$. AMP on the stacked fit, which fails at high coverage, and biwhitened AMP on the
product state, which returns no state, are not shown.}
\label{fig:matched}
\end{figure}

\begin{table}[!tb]
\centering
\caption{Median RGD infidelity relative to the observable design at $93\%$ coverage ($k=500$),
$\Nsh=2048$, over $50$ trials. The independent design and ML are the references of
Section~\ref{ss:matched}. Bootstrap $95\%$ intervals lie within $10\%$ of each entry.}
\label{tab:estimators}
\renewcommand{\arraystretch}{1.15}
\begin{tabular}{lcccccc}
\toprule
State & $\bar D$ & pooled & stacked & biwhitened & independent & ML \\
\midrule
GHZ         & $2.90$ & $0.66$ & $0.32$ & $0.37$ & $0.14$ & $0.07$ \\
W           & $3.28$ & $0.81$ & $0.38$ & $0.38$ & $0.17$ & $0.09$ \\
Haar-random & $1.92$ & $0.50$ & $0.23$ & $0.22$ & $0.13$ & $0.12$ \\
product     & $5.62$ & $1.30$ & $0.61$ & $0.12$ & $0.25$ & $0.12$ \\
\bottomrule
\end{tabular}
\end{table}

\subsection{Reconstruction at matched observable count}\label{ss:matched-results}

Figure~\ref{fig:matched} compares the designs at $\Nsh=2048$. With pooling, the median infidelity
of RGD is $1.2$ to $2.2$ times that of the observable design at $25\%$ coverage. The repeated
estimates help as coverage grows, and at $93\%$ the ratio is $0.66$, $0.81$ and $0.50$
for the GHZ, W and Haar-random states. Stacking roughly halves these ratios, to $0.32$, $0.38$ and
$0.23$, and biwhitening gives $0.37$, $0.38$ and $0.22$. For these three states the gain of
biwhitening over pooling is therefore the multiplicity weighting. At every coverage of $50\%$ and
above, biwhitened RGD beats pooled RGD in at least $48$ of $50$ paired trials. For the product state, whose
outcome distribution is the same in every setting, biwhitening also removes the correlation, and its
ratio at $93\%$ coverage is $0.12$, against $1.30$ with pooling and $0.61$ with stacking.

Table~\ref{tab:estimators} separates the multiplicity from the correlation. The independent design
has the multiplicity of the setting design without its correlation, so the ratio between the stacked
and independent columns is what the correlation costs a least-squares fit. The paired median of this
ratio is $1.7$ to $2.4$, and it exceeds one in all $50$ trials for every state. An estimator that
models the correlation does not pay this cost: ML on the setting data matches the independent design
for the Haar-random state and is below it for the other three. Its gain over stacking is $1.7$ to
$1.9$ for the Haar-random state, with $\bar D=1.92$, and $3$ to $5$ for the GHZ, W and product
states, with $\bar D\ge2.90$, at every coverage and shot count. This is the reading of
Section~\ref{ss:spectrum}: $D-1$ measures how far the noise of a setting is from white, which is what
a least-squares fit ignores. The mean $\bar D$ separates the unstructured state from the structured
ones, but it does not order the three structured states.

AMP at $\alpha=3$ agrees with RGD on the same data. The geometric mean of the per-trial ratio of
their infidelities is within $4\%$ of one for every design, estimator and state, with two exceptions.
Biwhitened AMP returns no state for the product state at any threshold we tried
($\alpha\in\{2,3,4\}$). For this state the separable weights reproduce full whitening, and AMP
fails here as it does under full whitening (Section~\ref{ss:matched}). AMP on the stacked fit fails in most trials for the GHZ and W states at $93\%$ coverage, so
the rescaling of biwhitening is what makes AMP usable on histograms. The threshold matters. At the
default $\alpha=2$, biwhitened AMP on the W state has $2.2$ times the error of RGD at $93\%$
coverage, while $\alpha=4$ removes the signal at $25\%$ coverage in up to $64\%$ of the trials with
the observable design.

\subsection{Reconstruction at matched cost}\label{ss:runs}

\begin{figure}[!tb]
\centering
\includegraphics[width=\linewidth]{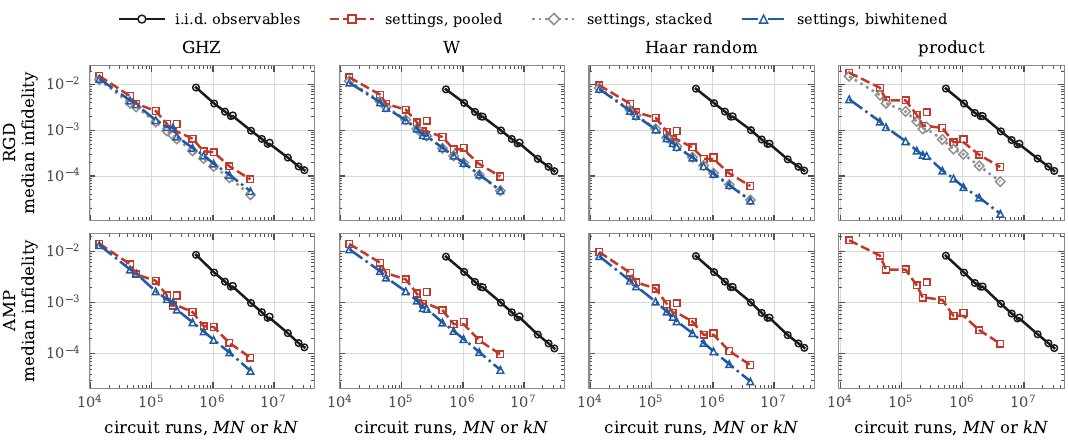}
\caption{Median infidelity against total circuit runs at $n=6$: $\Mobs\Nsh$ for the observable
design of $\Mobs$ observables and $k\Nsh$ for the setting design of $k$ settings, with $\Nsh$ shots
per measurement. Markers are individual configurations, and lines give
the best error reachable at or below each budget. The horizontal offset between the designs is
$\Mobs/k=(2^n-1)/\cbar$. Top row RGD, bottom row AMP with $\alpha=3$, without the two cases omitted
from Figure~\ref{fig:matched}. The setting design reaches every accuracy shown with fewer runs than
the observable design. The savings read from this figure are quantized by the grid of budgets.}
\label{fig:runs}
\end{figure}

Matched observable count isolates the effect of the sampling rule, but the cost of an experiment
is its number of circuit runs. By \eqref{eq:tradeoff}, the setting design is cheaper by $\Mobs/k=(2^n-1)/\cbar$, which
is $38.2$, $23.8$, $13.6$ and $7.6$ at coverage fractions $0.25$, $0.50$, $0.75$ and $0.93$, respectively.
Figure~\ref{fig:runs} places every configuration at its true cost. To reach a median infidelity of
$3\times10^{-4}$, RGD needs $1.675\times10^{7}$ runs with the observable design for every state.
With the setting design and pooling it needs $1.851\times10^{6}$ runs for the GHZ, W and product
states and $7.045\times10^{5}$ for the Haar-random state. Biwhitening lowers these to
$7.045\times10^{5}$ for the GHZ and W states, $4.628\times10^{5}$ for the Haar-random state and
$2.212\times10^{5}$ for the product state. The saving from settings is therefore $9.0$ to $23.8$
times with pooling and $23.8$ to $75.7$ times with biwhitening. Stacking matches biwhitening for the
GHZ, W and Haar-random states. AMP needs the same numbers of runs as RGD, apart from the two
exceptions of Section~\ref{ss:matched-results}.

A fixed budget of runs can also be split between more settings and more shots per setting. With
$2^{20}$ runs and $\Nsh=2^{20}/k$, the error of biwhitened RGD changes by less than $10\%$ as $k$
goes from $27$ to all $729$ settings, for every state, and it is $20$ to $66$ times below that of
the observable design with the same runs and the matched number of observables. With pooling, the
error grows by a factor of $1.9$ to $2.7$ over the same range of $k$. A histogram fit thus makes the number of settings
nearly irrelevant at this size, while pooling favors few settings.

\subsection{Readout error}\label{ss:readout}

Readout error flips each outcome bit independently with probability $\varepsilon$. Since the flip
probability is the same in every basis, this is a local depolarizing channel applied before the
measurement, which multiplies each expectation $\langle P\rangle$ by $(1-2\varepsilon)^{|P|}$ and
makes the measured state mixed. We added it at $\varepsilon=0.01$ and $0.03$, at $93\%$ coverage and
$\Nsh=2048$, and reconstructed without modeling it. The rank-one fits keep their advantage. Relative to the
observable design under the same error, biwhitened RGD reaches $0.33$, $0.53$, $0.19$ and $0.17$
for the GHZ, W, Haar-random and product states at $\varepsilon=0.03$, against $0.37$, $0.38$,
$0.22$ and $0.12$ without error. ML, whose weights come from the model probabilities, does not
survive the mismatch: at $\varepsilon=0.01$, its error is $54$ and $19$ times that of the observable
design for the GHZ and W states. AMP does not constrain the rank, and with biwhitening it keeps part
of the mixing that the error adds to the data. At $\varepsilon=0.01$ its infidelity is $12$ and
$30$ times that of RGD for the GHZ and W states, while pooled AMP stays within $3\%$ of RGD.

\subsection{A check at seven qubits}\label{ss:n7}

We repeated the comparison at $n=7$ with $\Nsh\in\{1024,4096\}$, the same coverage fractions and $20$
trials per configuration. Here $\Mobs/k$ is $69.8$, $41.2$, $22.0$ and $11.4$ at coverage fractions
$0.25$, $0.50$, $0.75$ and $0.93$. RGD behaves as it does at $n=6$. At $93\%$ coverage, its median
infidelity relative to the observable design is $0.70$ to $0.83$ with pooling and $0.34$ to $0.38$
with biwhitening for the GHZ and W states, and $0.43$ to $0.45$ and $0.18$ for the Haar-random
state. To reach a median infidelity of $3\times10^{-4}$, RGD needs $6.24\times10^{7}$ runs with the
observable design, $11.4$ to $27.3$ times fewer with pooling and $27.3$ to $45.5$ times fewer with
biwhitening.

AMP needs a larger threshold at $n=7$. With biwhitening and $\alpha=2$, it returns a state of
infidelity above $0.1$ at $93\%$ coverage in every trial for the Haar-random state and in $75\%$ of
the trials for the GHZ state, and at $75\%$ coverage in $50\%$ and $35\%$. With $\alpha=4$, no
trial fails at coverage fractions of $0.50$ and above, and AMP is within $3\%$ of RGD on the same
data. The threshold assumes a measurement map that senses every direction of the state equally on
average, as observables chosen uniformly at random do. The biwhitened map weights each observable
roughly by the number of settings that cover it, from one for an observable of weight $n$ to about
$k/3$ for an observable of weight one, and this spread grows with $n$, so the threshold has to grow
with it. We leave a rule for choosing $\alpha$, and AMP variants for general measurement maps such
as vector AMP \cite{rangan2019vector}, to future work. The reconstruction results at $n=7$ are
those of RGD.

\section{Conclusion}\label{sec:conclusion}

We summarize the covering results and the design effect, and list the questions
they leave open.
Random measurement settings cover the Pauli observables with a sharp threshold at the growth base
$3^{3/4}$. This base is set by the mean weight of an observable, and its natural logarithm is the rate--distortion
function of covering at zero distortion. Theorem~\ref{thm:kdelta} gives the number of
settings for a $1-\delta$ fraction explicitly.  Given $n$ and a target coverage, \eqref{eq:kstar-rand} tells an experimenter how
many settings to schedule. No selection rule changes the rate or the $\sqrt n$ term, and for large
$n$, random sampling needs fewer than $1.69$ times as many settings as the best rule.

Coverage does not determine what a setting provides. Its $2^n$ estimates come from one histogram
and are correlated. The design effect $D=2^n/\nueff$, where $\nueff=(\sum_bp(b)^2)^{-1}$ is the
effective support size of the outcome distribution, equals the squared Pauli expectation that the
setting captures. It is below $2$ on average over Haar-random states and settings, and its mean over
settings grows like $\tfrac12(4/3)^n$ for the GHZ and W states. The design effect can be estimated
without bias from the count histograms. In six-qubit simulations, the setting design
needs $7.6$ to $38$ times fewer circuit runs than the same number of independently measured
observables. Fitting the histograms instead of pooled estimates roughly halves the error at $93\%$
coverage, mainly because the fit weights each observable by the number of settings that cover it.
The correlation costs a least-squares fit about a factor of two, an estimator that models it does
not pay this cost, and the gain from modeling it is larger for states with a larger mean design
effect. Together, these let the setting design reach infidelity $3\times10^{-4}$ with $24$ to $76$
times fewer circuit runs than independent observables.

Several questions remain open. Compressed-sensing tomography recovers a rank-$r$ state on $n$
qubits from $O(r\,2^nn^2)$ randomly chosen Pauli observables, each measured separately
\cite{gross2010compressed}. A setting provides $2^n$ Pauli expectations at once, so combining
settings with compressed-sensing recovery could reduce the number of settings to
$O(r\,\mathrm{poly}(n))$ for typical low-rank states. We have not shown this, and the correlation within a setting, or a
state such as GHZ whose information is concentrated in a few settings, may raise the count.

Corollary~\ref{cor:constant} bounds how much any selection rule can gain over random sampling, but
the constant for the best rule is not known. For bounded weight, random sampling attains the rate,
but we give no deterministic construction that does. A derandomized
construction in the spirit of Huang, Kueng and Preskill \cite{huang2021efficient} is a natural
candidate. Finally, two settings
that cover the same number of observables can differ by a factor $2^{n-1}$ in $\nueff$, and
\eqref{eq:collision-est} makes the design effect estimable during an experiment. Adaptive setting selection
\cite{garciaperez2021learning,shlosberg2023adaptive}, locally biased shadows
\cite{hadfield2022measurements} and ShadowGrouping \cite{gresch2025guaranteed} are the closest
existing methods, but they select settings for prescribed observables rather than for the
state being measured. Whether the design effect can guide the choice of settings is open, and we
leave adaptive protocols to future work.

At the same time, our study has some limitations that future work could address.
The simulations use $n=6$ and $n=7$, rank-one reconstruction, the GHZ, W, Haar-random and product
states, and $50$ trials per configuration at $n=6$ and $20$ at $n=7$. They demonstrate the
mechanism, not a validated method. The AMP threshold must be tuned for the setting design and grows
with $n$ (Section~\ref{ss:n7}), and biwhitened AMP fails on the product state
(Section~\ref{ss:matched-results}).
Corollary~\ref{cor:constant} compares random sampling with the counting bound, which is a lower
bound on the optimum, and its constant $1.69$ holds in the limit of large $n$. At $n\le20$, sampling without replacement lowers $k^\star$ by an amount comparable to the
$n^{-1/2}$ term of Theorem~\ref{thm:kdelta}, so \eqref{eq:Ecov} should be evaluated directly
(Section~\ref{ss:cost}).

\section*{Author contributions}
KC and NS developed the ideas. KC performed the simulations and wrote the manuscript. NS, BNB and
DB edited the manuscript, and BNB and DB reviewed it and contributed through discussions.
The authors acknowledge that AI tools, specifically Claude Opus 5.0 and 5.5, were used to help with brainstorming ideas, generating proofs,
writing code, formatting the paper, and performing literature surveys. All theorems,
proofs, numerical results, and references were independently checked by the
authors, who assume responsibility for all content.

\section*{Acknowledgments}
KC and DB were supported in part by the NSF Institute for Robust Quantum Simulation (RQS), under award number CON-00002123-Z2253611.
BNB was supported by the U.S. Department of Energy, Advanced Scientific Computing Research, under contract number DE-SC0025384.

\bibliographystyle{quantum}
\bibliography{refs}

\appendix
\section{Analytical preliminaries}\label{app:prelim}
\begin{stdresult}[Elementary inequalities]\label{std:elem}
For $x\in[0,1]$ and real $k\ge1$, we have
\[
   (1-x)^k\le e^{-kx},\qquad (1-x)^k\ge1-kx,\qquad 1-(1-x)^k\le kx .
\]
The first holds because $1-x\le e^{-x}$, the second is
Bernoulli's inequality, which holds for every real exponent $k\ge1$, and the third is Bernoulli's
inequality rearranged.
\end{stdresult}

\begin{stdresult}[Chernoff bound]\label{std:chernoff}
For $W\sim\Bin(n,p)$ and $a>p$,  we have $\Prob[W\ge an]\le e^{-n\KL(a\|p)}$, where $\KL(a\|p)$ is defined in \eqref{eq:dap}. Symmetrically,
$\Prob[W\le an]\le e^{-n\KL(a\|p)}$ for $a<p$. Since $\KL(a\|p)>0$ for $a\neq p$, any fixed deviation of
the weight fraction $W/n$ from $\tfrac34$ has probability at most $e^{-cn}$ for some $c>0$. This is the
method-of-types estimate behind typicality arguments in source coding \cite[Ch.~11]{cover2006elements}.
\end{stdresult}

\begin{stdresult}[Berry--Esseen and the local limit theorem]\label{std:berry}
For the standardized binomial distribution, we have
\[\sup_x\big|\Prob[(W-\mu_W)/\sigma_W\le x]-\Phi(x)\big|=O(n^{-1/2}).\] Moreover,
$p_w=\sigma_W^{-1}\phi\bigl((w-\mu_W)/\sigma_W\bigr)(1+O(n^{-1/2}))$ uniformly in the bulk, so the
largest $p_w$ is $\Theta(n^{-1/2})$ and $p_w$ changes by a factor $1+O(n^{-1/2})$ across any window of
$O(1)$ consecutive weights. We write $f$ for the continuous Edgeworth density of $W$, the Gaussian density with the
corrections of the Edgeworth expansion through order $n^{-1}$
\cite{feller1971probability,petrov1975sums},
\[
   f(w):=\frac1{\sigma_W}\,\phi(x)\Bigl[1+\frac{\gamma_1}{6}\He_3(x)+\frac{\gamma_2}{24}\He_4(x)
      +\frac{\gamma_1^2}{72}\He_6(x)\Bigr],\qquad x=\frac{w-\mu_W}{\sigma_W}.
\]
Here $\He_3(x)=x^3-3x$, $\He_4(x)=x^4-6x^2+3$ and $\He_6(x)=x^6-15x^4+45x^2-15$ are Hermite
polynomials, and $\gamma_1=(1-2p)/\sqrt{np(1-p)}$ and $\gamma_2=(1-6p(1-p))/(np(1-p))$ are the
skewness and excess kurtosis of $\Bin(n,p)$, here with $p=\tfrac34$. Its derivatives satisfy
$f^{(j)}=O(\sigma_W^{-1-j})$. The weight $W$ takes integer values, its lattice has span one, and it
has moments of every order, so the local Edgeworth expansion for lattice distributions
\cite[Ch.~VII]{petrov1975sums} gives
\begin{equation}\label{eq:lattice-edgeworth}
   p_w=f(w)+O(n^{-2})\quad\text{uniformly in }w\in\mathbb Z .
\end{equation}
\end{stdresult}

\begin{stdresult}[Quantile conversion]\label{std:quantile}
For fixed $\delta\in(0,1)$, the density of $W$ at its $(1-\delta)$-quantile
is $\Theta(n^{-1/2})$ by \stdref{std:berry}, so moving a threshold by one unit of weight changes the tail
probability by $\Theta(n^{-1/2})$. A change of $O(n^{-1/2})$ in a tail probability therefore moves the
quantile by $O(1)$. This needs $\delta$ fixed, which keeps the quantile in the bulk.
\end{stdresult}

\begin{stdresult}[Cornish--Fisher expansion]\label{std:cf}
With $z=\Phi^{-1}(1-\delta)$, the $(1-\delta)$-quantile of the Edgeworth density $f$ satisfies
\cite{cornish1938moments,petrov1975sums}
\[
   \widetilde Q_{1-\delta}=\mu_W+\sigma_W\Bigl[z+\frac{\gamma_1}6(z^2-1)+\frac{\gamma_2}{24}(z^3-3z)
      -\frac{\gamma_1^2}{36}(2z^3-5z)\Bigr]+O(n^{-1}).
\]
For $p=\tfrac34$, we have $\sigma_W\gamma_1=-\tfrac12$, $\sigma_W\gamma_2=-\sqrt3/(6\sqrt n)$ and
$\sigma_W\gamma_1^2=\sqrt3/(3\sqrt n)$. The skewness correction is therefore $-(z^2-1)/12$, and the
two corrections of order $n^{-1/2}$ add to $-\sqrt3\,(11z^3-29z)/(432\sqrt n)$.
\end{stdresult}

\begin{stdresult}[Poisson summation]\label{std:poisson}
For a suitable function $g$ \cite{stein2003fourier}, we have
\[
   \sum_{w\in\mathbb Z}g(w)=\sum_{\ell\in\mathbb Z}\widehat g(2\pi\ell),\qquad
   \widehat g(\xi)=\int_{\mathbb R} g(w)e^{-i\xi w}\,dw.
\] The term $\ell=0$ is $\int g$, and
the terms $\ell=\pm1$ contribute $2\operatorname{Re}\widehat g(2\pi)$, an oscillation of period one in
any translation of $g$. A smooth function $g$ has rapidly decaying $\widehat g$ and hence a small oscillation.
A function $g$ with a kink, that is, a jump in $g'$, has $\widehat g(2\pi\ell)$ decaying only like
$1/\ell^2$.
\end{stdresult}

\begin{stdresult}[Gamma function]\label{std:gamma}
The Gamma function is \[\Gamma(z)=\int_0^\infty u^{z-1}e^{-u}\,du,\]
and differentiating at $z=1$ gives $\int_0^\infty\ln u\,e^{-u}\,du=\Gamma'(1)=-\gamma$. Differentiating twice gives $\int_0^\infty\ln^2u\,e^{-u}\,du=\Gamma''(1)=\gamma^2+\pi^2/6$. From $\Gamma(1+it)=it\,\Gamma(it)$ and the
reflection formula, $|\Gamma(1+it)|^2=\pi t/\sinh(\pi t)$. At $t=2\pi/\ln3\approx5.7192$, this gives
$|\Gamma(1+2\pi i/\ln3)|\approx7.519\times10^{-4}$, and at $t=4\pi/\ln3$, it gives
$|\Gamma(1+4\pi i/\ln3)|\approx1.334\times10^{-7}$.
\end{stdresult}

\begin{stdresult}[Euler--Maclaurin formula]\label{std:em}
Let $g$ be a continuous, piecewise continuously differentiable and integrable function on
$\mathbb R$, with $g'$ integrable and $g(x)\to0$ as $|x|\to\infty$. Integrating by parts on each
interval $[j,j+1]$ against the periodic Bernoulli function $\widetilde B_1(x)=\{x\}-\tfrac12$, where
$\{x\}$ is the fractional part, gives \cite{apostol1999elementary}
\begin{equation}\label{eq:EM}
   \sum_{w\in\mathbb Z}g(w)-\int_{\mathbb R}g(x)\,dx=\int_{\mathbb R}\widetilde B_1(x)\,g'(x)\,dx .
\end{equation}
If $\rho$ is integrable and of bounded variation, a second integration by parts, against
$\tfrac12\widetilde B_2(x)=\tfrac12(\{x\}^2-\{x\}+\tfrac16)$, whose derivative is $\widetilde B_1$
and whose absolute value is at most $\tfrac1{12}$, gives
$\bigl|\int_{\mathbb R}\widetilde B_1\rho\bigr|\le\tfrac1{12}\operatorname{TV}(\rho)$, where
$\operatorname{TV}$ denotes total variation.
\end{stdresult}

\section{Proof of Theorem~\ref{thm:kdelta}}\label{app:thm2}

\subsection{The step approximation}
Write $\pmax(n):=\max_wp_w=\Theta(n^{-1/2})$, by the local limit theorem (\stdref{std:berry}).

\begin{lemma}[Step approximation]\label{lem:step}
Let $t\in\mathbb R$ and $k=3^{t}$. Then
\begin{align}
   \sum_{w\le t}p_w\,(1-3^{-w})^{k}&\ \le\ C_1\,\pmax(n),\label{eq:stepI}\\
   \sum_{w>t}p_w\bigl(1-(1-3^{-w})^{k}\bigr)&\ \le\ \tfrac32\,\pmax(n),\label{eq:stepII}
\end{align}
where $C_1:=\sum_{j\ge0}e^{-3^{j}}<0.42$ and the sums range over integers $w\in\{0,\dots,n\}$.
Consequently, \[\big|U(k)-\Prob[W>t]\big|\le2\,\pmax(n)=O(n^{-1/2}).\]
\end{lemma}

\begin{proof}
The difference $U(k)-\Prob[W>t]$ is the left side of \eqref{eq:stepI} minus the left side of
\eqref{eq:stepII}, so the last statement follows from $C_1+\tfrac32<2$. For \eqref{eq:stepI}, the elementary inequalities (\stdref{std:elem})
give $(1-3^{-w})^{k}\le e^{-3^{t-w}}$. Writing $w=\lfloor t\rfloor-j$ with $j\ge0$ implies
$t-w=\{t\}+j\ge j$, so $e^{-3^{t-w}}\le e^{-3^{j}}$, and bounding each $p_w$ by $\pmax(n)$, we get
\eqref{eq:stepI} with $C_1=0.4178\ldots$. For \eqref{eq:stepII}, the elementary inequalities (\stdref{std:elem}) give
$1-(1-3^{-w})^{k}\le3^{t-w}$. Writing $w=\lfloor t\rfloor+j$ with $j\ge1$, we have
\begin{equation}\label{eq:fracpart}
   \sum_{w>t}3^{t-w}=3^{\{t\}}\sum_{j\ge1}3^{-j}=\frac{3^{\{t\}}}{2}<\frac32,
\end{equation}
and bounding each $p_w$ by $\pmax(n)$ completes the proof.
\end{proof}

\subsection{The \texorpdfstring{$O(1)$}{O(1)} form}
By Lemma~\ref{lem:exp}, $\underline U\le\widetilde U\le U$. Near the crossover, $\underline U$ and
$U$ differ by a factor $1+O(k3^{-n})$, which is exponentially small in $n$ because $k3^{-n}$ is.
Below the crossover, both are doubly exponentially small. Since the density of $W$ near the root is
$\Theta(n^{-1/2})$, this moves the root of $\widetilde U(k)=\delta$ by $o(n^{-1/2})$ in $\log_3k$.
Therefore, it suffices to solve $U(k)=\delta$. Let $\wstar=\log_3k^\star$, taken as real-valued. Lemma~\ref{lem:step} at $t=\wstar$ gives
$\Prob[W>\wstar]=\delta+O(n^{-1/2})$, and quantile conversion (\stdref{std:quantile}) gives $\wstar=Q_{1-\delta}(W)+O(1)$, where
$Q_{1-\delta}(W):=\inf\{x:\Prob[W>x]\le\delta\}$. From the Berry--Esseen bound (\stdref{std:berry}), converted in the same
way, we obtain $Q_{1-\delta}(W)=\mu_W+\sigma_W\zd+O(1)$, and \eqref{eq:musigma} yields
$\log_3k^\star=\tfrac34n+\tfrac{\sqrt{3n}}4\zd+O(1)$.

\subsection{The explicit constants}
Define the miss profile $m(s):=e^{-3^{-s}}$, which increases from $0$ to $1$. For every $w\ge1$,
\[(1-3^{-w})^k=m(w-\wstar)\bigl(1+O(k3^{-2w})\bigr).\] When $w\ge\wstar-O(\log n)$, the error
$k3^{-2w}=3^{\wstar-2w}$ is exponentially small in $n$. Below that range, both sides are doubly
exponentially small. Hence,
\begin{equation}\label{eq:sumtoint}
   U(k)=\sum_{w\in\mathbb Z}p_w\,m(w-\wstar)+o(n^{-1})
      =\int_{-\infty}^\infty f(w)\,m(w-\wstar)\,dw+E_n+o(n^{-1}),
\end{equation}
where $f$ is the Edgeworth density from \stdref{std:berry} and
\[E_n:=\sum_{w\in\mathbb Z}f(w)m(w-\wstar)-\int_{-\infty}^\infty f(w)m(w-\wstar)\,dw\] is the
lattice error, calculated in Appendix~\ref{app:lattice}. The second equality replaces $p_w$ by
$f(w)$, which by \eqref{eq:lattice-edgeworth} changes each term by $O(n^{-2})$. The
$O(\sqrt{n\ln n})$ weights within $3\sigma_W\sqrt{\ln n}$ of $\mu_W$ therefore contribute
$O(n^{-3/2}\sqrt{\ln n})$. Outside this range, Hoeffding's inequality bounds the total of $p_w$ and
the Gaussian factor bounds the total of $|f|$, and both totals are $o(n^{-1})$.

The profile $m$ is the distribution function of the random variable $S:=G/\ln3$, where $G$ has the
standard Gumbel distribution, $\Prob[G\le x]=\exp(-e^{-x})$. Writing
$m(w-\wstar)=\int\mathbf 1\{s\le w-\wstar\}\,m'(s)\,ds$ and exchanging the order of integration
gives
\begin{equation}\label{eq:conv}
   \int_{-\infty}^\infty f(w)\,m(w-\wstar)\,dw=\int_{-\infty}^{\infty}m'(s)\,\bar F(\wstar+s)\,ds
      =\E\bigl[\bar F(\wstar+S)\bigr],\qquad \bar F(x):=\int_x^\infty f(s)\,ds.
\end{equation}
The substitution $u=3^{-s}$ gives the first two moments of $S$. It sends $s\to-\infty$ to
$u\to\infty$ and $s\to\infty$ to $u\to0$, and gives $s=-\ln u/\ln3$ and $m'(s)\,ds=-e^{-u}\,du$.
The reversal of orientation and the sign of $-e^{-u}\,du$ cancel, so
\begin{equation}\label{eq:mum}
   \mu_m:=\E[S]=\int_{u=\infty}^{u=0}\Bigl(-\frac{\ln u}{\ln3}\Bigr)\bigl(-e^{-u}\,du\bigr)
        =-\frac1{\ln3}\int_{0}^{\infty}\ln u\;e^{-u}\,du=\frac{\gamma}{\ln3}
\end{equation}
by the Gamma-function identity $\Gamma'(1)=-\gamma$ (\stdref{std:gamma}). In the same way,
$\E[S^2]=\Gamma''(1)/\ln^23$, and $\Gamma''(1)=\gamma^2+\pi^2/6$ gives
\begin{equation}\label{eq:vm}
   v_m:=\operatorname{Var}[S]=\frac{\pi^2}{6\ln^23}\approx1.363 .
\end{equation}
These are the mean $\gamma$ and variance $\pi^2/6$ of the standard Gumbel distribution, divided by
$\ln3$ and $\ln^23$.

Let $y:=\wstar+\mu_m$. Expanding $\bar F(\wstar+S)$ about $y$ to second order, with $\bar F'=-f$,
the first-order term has mean zero. The remainder is at most
$\tfrac16\sup|f''|\,|S-\mu_m|^3$, where $\sup|f''|=O(n^{-3/2})$ and $S$ has a finite third absolute
moment, so
\begin{equation}\label{eq:taylor}
   \E\bigl[\bar F(\wstar+S)\bigr]=\bar F(y)-\tfrac12v_m\,f'(y)+O(n^{-3/2}).
\end{equation}
By the $O(1)$ form, $(y-\mu_W)/\sigma_W=\zd+O(n^{-1/2})$, so differentiating $f$ gives
\[
   f'(y)=-\frac{\zd}{\sigma_W}\,f(y)\bigl(1+O(n^{-1/2})\bigr).
\]
Combining \eqref{eq:sumtoint}, \eqref{eq:conv} and \eqref{eq:taylor} with $U(k^\star)=\delta$
gives
\[
   \bar F(y)+\frac{\zd v_m}{2\sigma_W}\,f(y)=\delta-E_n+o(n^{-1}).
\]
Let $\widetilde Q_{1-\delta}$ solve $\bar F(x)=\delta$ and write $y=\widetilde Q_{1-\delta}+\Delta$.
Expanding $\bar F$ to first order, with $\bar F'=-f$ and $f'=O(f/\sigma_W)$, and dividing by
$f(y)=\Theta(n^{-1/2})$, we obtain
\begin{equation}\label{eq:Delta}
   \Delta=\frac{\zd v_m}{2\sigma_W}+\lambda_n+o(n^{-1/2}),\qquad \lambda_n:=\frac{E_n}{f(y)},
\end{equation}
so that $\wstar=\widetilde Q_{1-\delta}-\gamma/\ln3+\zd v_m/(2\sigma_W)+\lambda_n+o(n^{-1/2})$.
By \eqref{eq:musigma}, $\sigma_W=\sqrt{3n}/4$, so $\zd v_m/(2\sigma_W)=\pi^2\zd/(3\sqrt3\ln^23\sqrt n)$.
The Cornish--Fisher expansion (\stdref{std:cf}) implies
\[
   \widetilde Q_{1-\delta}=\tfrac34n+\tfrac{\sqrt{3n}}4\zd-\frac{\zd^{2}-1}{12}
      -\frac{\sqrt3\,(11\zd^3-29\zd)}{432\sqrt n}+O(n^{-1}),
\]
and the two terms of order $n^{-1/2}$ add to $c(\delta)/\sqrt n$ of \eqref{eq:cdelta}.
Appendix~\ref{app:lattice} shows that the $\limsup$ of $|E_n|/f(\wstar)$ is at most
$\epslat+3\times10^{-8}$. Since $f(y)/f(\wstar)\to1$, the same bound holds for $\lambda_n$, which
proves Theorem~\ref{thm:kdelta}.

The argument uses only that the profile is the distribution function of a random variable with a
finite third absolute moment. That variable shifts the crossover by its mean and widens the weight
distribution by its variance. To the $o(1)$ accuracy of Corollary~\ref{cor:constant}, only the mean
matters, and Appendix~\ref{app:thm3} applies the argument to a second profile.

\subsection{The lattice term}\label{app:lattice}
The lattice error $E_n$ is a classical fluctuation. A doubly exponential profile
$e^{-q^{-s}}$ summed over a lattice is a harmonic sum, and Mellin asymptotics gives its fluctuation as
a Fourier series in $\Gamma(1+2\pi i\ell/\ln q)$, following Flajolet, Gourdon and Dumas
\cite{flajolet1995harmonic}.  Our $\epslat$ is that
amplitude for $q=3$.

Write $t=\wstar$ and $g(w):=f(w)\,m(w-t)$. The function $g$ is smooth and integrable, so by Poisson summation (\stdref{std:poisson}), we have
\[
   E_n=\sum_{\ell\neq0}\widehat g(2\pi\ell)=2\operatorname{Re}\widehat g(2\pi)+\sum_{|\ell|\ge2}\widehat g(2\pi\ell).
\]
Integration by parts gives
\[\widehat g(2\pi)=\frac1{2\pi i}\int_{-\infty}^\infty\Bigl(f'(w)m(w-t)+f(w)m'(w-t)\Bigr)e^{-2\pi iw}\,dw.\] The term containing
$f'$ is smaller than the term containing $m'$ by a factor $O(1/\sigma_W)$, since $f'=O(f/\sigma_W)$. In
the term containing $m'$, $f$ can be evaluated at $t$ up to a factor $1+O(n^{-1/2})$, and substituting
$w=t+s$ gives
\begin{equation}\label{eq:ghat}
   \widehat g(2\pi)=\frac{f(t)\,e^{-2\pi it}}{2\pi i}\,\widehat{m'}(2\pi)\bigl(1+O(n^{-1/2})\bigr).
\end{equation}
With $u=3^{-s}$, so that $e^{-i\xi s}=u^{i\xi/\ln3}$, the Fourier transform of $m'$ is a Gamma
function,
\begin{equation}\label{eq:mhat}
   \widehat{m'}(\xi)=\int_{-\infty}^{\infty}m'(s)e^{-i\xi s}\,ds
      =\int_0^\infty e^{-u}u^{i\xi/\ln3}\,du=\Gamma\Bigl(1+\frac{i\xi}{\ln3}\Bigr),
\end{equation}
which is the harmonic-sum identity of \cite{flajolet1995harmonic} in our notation. The terms $\ell=\pm1$
therefore contribute an oscillation of period one in $t$ with amplitude
\[2|\widehat g(2\pi)|=f(t)\frac{|\Gamma(1+2\pi i/\ln3)|}{\pi}\bigl(1+o(1)\bigr)=f(t)\,\epslat\bigl(1+o(1)\bigr),\] where
$\epslat=7.519\times10^{-4}/\pi=2.394\times10^{-4}$ by the Gamma-function values in \stdref{std:gamma}. The terms with $|\ell|\ge2$ are governed by
$|\Gamma(1+2\pi\ell i/\ln3)|$, which is $1.334\times10^{-7}$ at $\ell=2$ and decays like
$e^{-\pi^2\ell/\ln3}$. Their total contribution to the amplitude is below $3\times10^{-8}$,
and the largest part, $2.1\times10^{-8}$, comes from $\ell=\pm2$. Hence,
$\limsup|E_n|/f(\wstar)\le\epslat+3\times10^{-8}$.

\section{Proofs of Theorem~\ref{thm:lb} and Corollary~\ref{cor:constant}}\label{app:thm3}

\subsection{Proof of Theorem~\ref{thm:lb}}
The bound \eqref{eq:anybound} was proved in the proof of Theorem~\ref{thm:threshold}. Dividing by $4^n$ and using
\eqref{eq:weightdist}, with $\wstar=\log_3k$, we obtain
\begin{equation}\label{eq:optprofile}
   \frac{\Ncov(\mathcal S)}{4^n}\ \le\ \sum_{w\in\mathbb Z}p_w\min\bigl(1,3^{\wstar-w}\bigr).
\end{equation}
  At $k=k^\star(n,\delta)$, the random sample satisfies $\E[\Ncov(k)]\ge(1-\delta)4^n$, so
some realization covers a $1-\delta$ fraction, and $k^\star_{\mathrm{opt}}\le k^\star$. Theorem~\ref{thm:kdelta}
then proves the upper bound in \eqref{eq:kopt}. For the lower bound, suppose some collection of $k$
settings covers a $1-\delta$ fraction, so the right side of \eqref{eq:optprofile} is at least
$1-\delta$. Bounding each $p_w$ by $\pmax(n)$ and using \eqref{eq:fracpart}, the part of that sum with
$w>\wstar$ is at most $\tfrac32\pmax(n)=O(n^{-1/2})$, and the part with $w\le\wstar$ is at most $\Prob[W\le\wstar]$. Hence,
$\Prob[W>\wstar]\le\delta+O(n^{-1/2})$, and quantile conversion (\stdref{std:quantile}) with the Berry--Esseen bound (\stdref{std:berry}) implies
$\wstar\ge\tfrac34n+\tfrac{\sqrt{3n}}4\zd-O(1)$.

\subsection{Proof of Corollary~\ref{cor:constant}}
Rounding $k$ to an integer changes $\log_3k$ by $O(1/k)$, and sampling with or without replacement
changes $\log_3k^\star_{\mathrm{rand}}$ by $o(1)$, so it suffices to compare the real roots
$t_{\mathrm{rand}}$ of $U(3^t)=\delta$ and $t_{\mathrm{cb}}$ of $V(t)=\delta$, where
\begin{equation}\label{eq:Vdef}
   V(t):=1-\sum_{w\in\mathbb Z}p_w\min\bigl(1,3^{t-w}\bigr)=\sum_{w\in\mathbb Z}p_w\,m_{\mathrm{cb}}(w-t),\qquad
   m_{\mathrm{cb}}(s):=\max\bigl(0,1-3^{-s}\bigr),
\end{equation}
is the smallest uncovered fraction that \eqref{eq:anybound} allows with $3^t$ settings.

\emph{The roots $t_{\mathrm{rand}}$ and $t_{\mathrm{cb}}$ up to the lattice term.} The profile $m_{\mathrm{cb}}$ is the distribution function of an exponential random variable with
rate $\ln3$, whose density is $m_{\mathrm{cb}}'(s)=\ln3\cdot3^{-s}\,\mathbf 1\{s>0\}$ and whose
mean is $\mu_{\mathrm{cb}}=\int_0^\infty s\ln3\cdot3^{-s}\,ds=1/\ln3$. The argument of
Appendix~\ref{app:thm2}, kept to order $o(1)$, therefore applies to $m$ and $m_{\mathrm{cb}}$ and
gives
\[
   t_{\mathrm{rand}}=\widetilde Q_{1-\delta}-\frac{\gamma}{\ln3}+\frac{E_n}{f(t_{\mathrm{rand}})}+o(1),
   \qquad
   t_{\mathrm{cb}}=\widetilde Q_{1-\delta}-\frac{1}{\ln3}+\frac{E_n^{\mathrm{cb}}}{f(t_{\mathrm{cb}})}+o(1),
\]
where \[E_n^{\mathrm{cb}}:=\sum_{w\in\mathbb Z}f(w)m_{\mathrm{cb}}(w-t_{\mathrm{cb}})-\int_{\mathbb R} f(w)m_{\mathrm{cb}}(w-t_{\mathrm{cb}})\,dw.\]
By subtracting, we get
\[
   t_{\mathrm{rand}}-t_{\mathrm{cb}}=\frac{1-\gamma}{\ln3}+\frac{E_n}{f(t_{\mathrm{rand}})}
   -\frac{E_n^{\mathrm{cb}}}{f(t_{\mathrm{cb}})}+o(1),
\]
and Appendix~\ref{app:lattice} gives $\limsup|E_n|/f(t_{\mathrm{rand}})\le\epslat+3\times10^{-8}$.

\emph{The lattice term for the counting bound.} Let $t=t_{\mathrm{cb}}$ and
$g(w):=f(w)\,m_{\mathrm{cb}}(w-t)$. The profile $m_{\mathrm{cb}}$ has a kink at $s=0$, so the
Poisson-summation terms $\widehat g(2\pi\ell)$ (\stdref{std:poisson}) decay only like $1/\ell^2$, and
we sum them in closed form instead. The function $g$ is continuous and piecewise smooth and tends
to zero at $\pm\infty$, so the Euler--Maclaurin formula \eqref{eq:EM} gives
\[
   E_n^{\mathrm{cb}}=\int_{\mathbb R}\widetilde B_1(w)\,g'(w)\,dw .
\]
We split $g'(w)=f(t)\,m_{\mathrm{cb}}'(w-t)+\rho(w)$, where
$\rho(w):=f'(w)\,m_{\mathrm{cb}}(w-t)+\bigl(f(w)-f(t)\bigr)\,m_{\mathrm{cb}}'(w-t)$. The first term
carries the kink, and it contributes $f(t)\,h(t)$, where
\begin{equation}\label{eq:hcb}
   h(t):=\int_0^\infty\widetilde B_1(t+s)\,\ln3\cdot3^{-s}\,ds
        =\{t\}-\frac12+\frac1{\ln3}-\frac{3^{\{t\}}}{2}.
\end{equation}
The closed form follows by reading $\ln3\cdot3^{-s}$ as an exponential density: its mean is
$1/\ln3$, and for $t\in[0,1)$ we have $\int_0^\infty\lfloor t+s\rfloor\,\ln3\cdot3^{-s}\,ds
=\sum_{j\ge1}3^{t-j}=3^t/2$. The Fourier coefficients of $h$ are the Poisson-summation terms
$\ln3/\bigl(2\pi i\ell\,(2\pi i\ell-\ln3)\bigr)$, $\ell\neq0$, which decay like $1/\ell^2$. The
function $h$ is continuous with period one, and its smallest value is $h(0)=1/\ln3-1\approx-0.090$.
The remainder $\rho$ is continuous, because the jump of $m_{\mathrm{cb}}'$ at $w=t$ is multiplied by
$f(t)-f(t)=0$. Its total variation is
$\operatorname{TV}(\rho)\le\int_{\mathbb R}|f''|+2\sup|f'|+\sup|f'|\int_0^\infty s\,|m_{\mathrm{cb}}''(s)|\,ds
=O(n^{-1})$, since $\sup|f'|=O(n^{-1})$ and $\int_{\mathbb R}|f''|=O(n^{-1})$. The second-order
bound in \stdref{std:em} therefore gives $\bigl|\int_{\mathbb R}\widetilde B_1\rho\bigr|=O(n^{-1})$,
and hence $E_n^{\mathrm{cb}}=f(t)\,h(t)+O(n^{-1})$. Because $f(t)=\Theta(n^{-1/2})$ in the bulk,
\[
   \frac{E_n^{\mathrm{cb}}}{f(t_{\mathrm{cb}})}=h(t_{\mathrm{cb}})+o(1)\ \ge\ \frac1{\ln3}-1+o(1).
\]

\emph{Result.} Combining the bounds on $E_n$ and $E_n^{\mathrm{cb}}$, we obtain
\[
   t_{\mathrm{rand}}-t_{\mathrm{cb}}\le\frac{1-\gamma}{\ln3}+\epslat+3\times10^{-8}
   -\Bigl(\frac1{\ln3}-1\Bigr)+o(1)=1-\frac{\gamma}{\ln3}+\epslat+3\times10^{-8}+o(1),
\]
which is \eqref{eq:constant}. Hence,
$k^\star_{\mathrm{rand}}/k^\star_{\mathrm{cb}}\le3^{1-\gamma/\ln3+\epslat+3\times10^{-8}}(1+o(1))
=3e^{-\gamma}\,3^{\epslat+3\times10^{-8}}(1+o(1))$, and
$3e^{-\gamma}\,3^{\epslat+3\times10^{-8}}=1.6848\ldots<1.69$. Every collection of settings obeys
\eqref{eq:anybound}, so $k^\star_{\mathrm{cb}}\le k^\star_{\mathrm{opt}}$, and the bound
$k^\star_{\mathrm{rand}}<1.69\,k^\star_{\mathrm{opt}}$ of Corollary~\ref{cor:constant} follows.

\section{Proof of Corollary~\ref{cor:bounded}}\label{app:cor}
Restrict to observables of weight at most $\omega n$, and let $W_\omega$ be the weight of one drawn
uniformly from them, so $\Prob[W_\omega=w]=p_w/\Prob[W\le\omega n]$ for $w\le\omega n$. Write
$w_0:=\lfloor\omega n\rfloor$ for the top weight. As in Appendix~\ref{app:thm2}, it suffices to work
with sampling with replacement, for which the uncovered fraction at $k$ settings is
$U_\omega(k):=\sum_{w\le w_0}\Prob[W_\omega=w]\,(1-3^{-w})^{k}$.

\emph{The truncated weight concentrates at the top weight.} Consecutive shell probabilities satisfy
\begin{equation}\label{eq:shellratio}
   \frac{p_{w+1}}{p_w}=\frac{3(n-w)}{w+1},
\end{equation}
which decreases in $w$ and equals $3(1-\omega)/\omega+O(1/n)>1$ at $w=\omega n$, since $\omega<\tfrac34$.
Hence, the ratio is at least a constant $\lambda=\lambda(\omega)>1$ for all $w\le w_0$, so
$p_{w_0-j}\le\lambda^{-j}p_{w_0}$ for every $j\ge0$, and
\begin{equation}\label{eq:edge-mass}
   \Prob[W_\omega=w_0]\ \ge\ 1-\lambda^{-1},\qquad
   \Prob[W_\omega\le w_0-j]\ \le\ \frac{\lambda^{-j}}{1-\lambda^{-1}}\quad\text{for every }j\ge0.
\end{equation}
Moreover, for every fixed $\eta>0$, the Chernoff bound (\stdref{std:chernoff}) and the estimate
$\Prob[W\le\omega n]\ge p_{w_0}=e^{-n\KL(\omega\|3/4)+O(\log n)}$ imply
\[
   \Prob[W_\omega\le(\omega-\eta)n]\le e^{-n\Delta_\eta+O(\log n)},\qquad
   \Delta_\eta:=\KL(\omega-\eta\,\|\,\tfrac34)-\KL(\omega\,\|\,\tfrac34)>0,
\] because $a\mapsto\KL(a\|\tfrac34)$ is
decreasing on $(0,\tfrac34)$.

\emph{Part (i).} The proof of Theorem~\ref{thm:threshold} applies with $W_\omega$ in place of $W$ and
$\omega$ in place of $\tfrac34$. If $\log_3\beta>\omega$, the crossover $\wstar=n\log_3\beta+o(n)$ lies
above every weight in the band, and the doubly exponential bound of part~(i) of that proof shows that
$U_\omega(k)\to0$. If $\log_3\beta<\omega$, the crossover lies a distance $\Theta(n)$ below $w_0$, the
weights above it carry $1-o(1)$ of the truncated probability by the concentration just shown, and the
argument of part~(ii) shows that they are missed.

\emph{Part (ii), upper bound.} Let $J:=\wstar-w_0$. For every $w\le w_0$, we have $k3^{-w}\ge3^{J}$, and by
the elementary inequalities (\stdref{std:elem}), we get $U_\omega(k)\le e^{-3^{J}}$. Hence, $U_\omega(k)\le\delta$ once $J\ge\log_3\ln(1/\delta)$, which
gives $\log_3k^\star_\omega\le w_0+\log_3\ln(1/\delta)+1$, the $1$ accounting for rounding $k$ to an
integer.

\emph{Part (ii), lower bound.} Keeping only the term $w=w_0$ and using \eqref{eq:edge-mass} and
$\ln(1-x)\ge-x-x^2$ for $x\in[0,\tfrac12]$, we obtain
\[
   U_\omega(k)\ \ge\ \Prob[W_\omega=w_0]\,(1-3^{-w_0})^{k}\ \ge\ (1-\lambda^{-1})\exp\bigl(-3^{J}(1+3^{-w_0})\bigr),
\]
which exceeds $\delta$ whenever $J<\log_3\ln((1-\lambda^{-1})/\delta)-o(1)$. When $\delta<1-\lambda^{-1}$,
this gives $\log_3k^\star_\omega\ge w_0+\log_3\ln(1/\delta)-O(1)$. For a general fixed
$\delta\in(0,1)$, keep the top $J_1+1$ weights instead: if $J\le-J_0$ with $J_0\ge J_1$, then
$k3^{-w}\le3^{J_1-J_0}$ for $w\ge w_0-J_1$. Equation \eqref{eq:edge-mass} and the elementary inequalities (\stdref{std:elem}) imply
\[
   U_\omega(k)\ \ge\ \Prob[W_\omega\ge w_0-J_1]\bigl(1-3^{J_1-J_0}\bigr)
   \ \ge\ \Bigl(1-\frac{\lambda^{-J_1-1}}{1-\lambda^{-1}}\Bigr)\bigl(1-3^{J_1-J_0}\bigr),
\]
which exceeds $\delta$ once $J_1$ and then $J_0$ are large enough, depending on $\delta$ and $\omega$
only. Hence, $\log_3k^\star_\omega\ge w_0-J_0=\omega n-O(1)$.

Since $w_0=\omega n+O(1)$, the upper and lower bounds yield $\log_3k^\star_\omega=\omega n+O(1)$ for fixed
$\delta$. If $\delta_n\to0$ with $\ln(1/\delta_n)=3^{o(n)}$, then $\delta_n<1-\lambda^{-1}$ for large
$n$ and $\log_3\ln((1-\lambda^{-1})/\delta_n)=\log_3\ln(1/\delta_n)+o(1)$, so the same bounds give
$\log_3k^\star_\omega=\omega n+\log_3\ln(1/\delta_n)+O(1)$ with the $O(1)$ depending on $\omega$ only.
The growth condition on $\delta_n$ keeps $k3^{-n}\to0$, which the transfer between sampling with and
without replacement requires.

\end{document}